\documentclass[aas2pp4,natbib209]{emulateapj}

  \slugcomment{2019 Mar 27}

\shorttitle{Polarization in G34}
\shortauthors{Tang et al.}

\begin{document}


%\title{Evidence of material channeled by magnetic field in the infrared dark cloud G34.43? A Comparison between Magnetic Field, Turbulence, and Gravity}
\title{Gravity, Magnetic Field, and Turbulence: Relative Importance and Impact on Fragmentation in the Infrared Dark Cloud G34.43+00.24}


\author{Ya-Wen Tang\altaffilmark{1}, Patrick M. Koch\altaffilmark{1}, Nicolas Peretto\altaffilmark{2}, Giles Novak\altaffilmark{3}, Ana Duarte-Cabral\altaffilmark{2},  Nicholas L. Chapman\altaffilmark{3}, Pei-Ying Hsieh\altaffilmark{1}, Hsi-Wei Yen\altaffilmark{1}}
\affil{1 Academia Sinica, Institute of Astronomy and Astrophysics, Taipei, Taiwan}
\affil{2 School of Physics \& Astronomy, Cardiff University, Queen's building, The Parade, Cardiff, CF24 3AA, UK}
\affil{3 Center for Interdisciplinary Exploration and Research in Astrophysics (CIERA) and Department of Physics \& Astronomy, Northwestern University, USA}
%\author{C. D. Biemesderfer\altaffilmark{4,5}}
%\affil{National Optical Astronomy Observatories, Tucson, AZ 85719}
%\email{aastex-help@aas.org}

%\and

%\author{R. J. Hanisch\altaffilmark{5}}
%\affil{Space Telescope Science Institute, Baltimore, MD 21218}


%\altaffiltext{1}{Visiting Astronomer, Cerro Tololo Inter-American Observatory.
%CTIO is operated by AURA, Inc.\ under contract to the National Science
%\altaffiltext{2}{Society of Fellows, Harvard University.}
%\altaffiltext{3}{present address: Center for Astrophysics,
%    60 Garden Street, Cambridge, MA 02138}
%\altaffiltext{4}{Visiting Programmer, Space Telescope Science Institute}
%\altaffiltext{5}{Patron, Alonso's Bar and Grill}


\begin{abstract}

We investigate the interplay between magnetic (B) field, gravity, and turbulence in the fragmentation process of cores within the filamentary infrared dark cloud G34.43+00.24. 
%
% method:
% 
%The B field morphology across G34.43+00.24 is traced with thermal dust polarization at 350 $\mu$m with an angular resolution of 10$\arcsec$ (0.18 pc).
%Local velocity gradients are derived from N$_{2}$H$^{+}$, tracing motion in the plane of sky, and are compared with the observed local B field orientations.
We observe the magnetic field (B) morphology across G34.43, traced with thermal dust polarization at 350 $\mu$m with an angular resolution
of 10$\arcsec$ (0.18 pc), and compare with the kinematics obtained from N$_{2}$H$^{+}$ across the filament. 
We derive local velocity gradients from N$_{2}$H$^{+}$, tracing motion in the plane of sky, and compare with the observed
local B field orientations in the plane of sky. 
The B field orientations are found to be perpendicular to the long axis of the filament toward the MM1 and MM2 ridge, suggesting that the B field can guide material toward the filament. 
Toward MM3, the B field orientations appear more parallel to the filament and aligned with the elongated MM3 core, hinting a different B field role.
% 
% results:
%
Besides a large-scale east-west velocity gradient, we find a close
alignment between local B field orientations and local velocity gradients toward the MM1/MM2 ridge.
This local correlation in alignment suggests that gas motions are influenced by the magnetic field morphology or vice versa.
Additionally, this alignment seems to be getting even closer with increasing integrated emission in N$_{2}$H$^{+}$, possibly indicating that a growing gravitational pull is more and more aligning B field and gas motion.
We analyze and quantify B field, gravity, turbulence, and their relative importance toward the MM1, MM2 and MM3 regions with various techniques over two scales, a larger clump area at 2 pc scale and the smaller core area at 0.6 pc scale. 
While gravitational energy, B field, and turbulent pressure 
all grow systematically from large to small scale, the ratios among the three constituents develop clearly differently over scale. 
We propose that this varying relative importance between B field, gravity, and turbulence over scale drives and explains the different fragmentation types seen at sub-pc scale (no fragmentation in MM1; aligned 
fragmentation in MM2; clustered fragmentation in MM3). 
We discuss uncertainties, subtleties, and the robustness of our conclusion, and we stress the 
need of a multi-scale joint analysis to understand the dynamics in these systems. 
\end{abstract}


\section{Introduction}
Recent {\it Herschel} results show that molecular clouds are mostly filamentary \citep{2013Andre}.
One mechanism to form filamentary structures is through compressive flows, where the filaments appear at the interfaces where flows collide \citep[for example,][]{2007Ballesteros-Paredes,2015Inutsuka}.
%and are compressed.
Molecular clouds are both turbulent and threaded by magnetic (B) fields which possibly explains the observed low star-formation rate \citep[e.g.,][and the references therein]{2015Enrique}. 
Depending on what the dominant force is in shaping a cloud, fragmentation and 
dense core formation will be different \citep[e.g.,][]{2011Hennebelle}.  
B fields are recognized as one of the key components in star-formation theories \citep[e.g.,][]{2007McKee}. 
Nevertheless, their exact role in the formation and evolution of molecular clouds is still a matter of debate in the literature. 
Finally, it has been suggested that local clouds often form hub-filament systems \citep{2009Myers}, and star clusters are formed in the hub.
In order to test and constrain the formation mechanism of filaments, observations of both magnetic field and gas kinematics within a filament are crucial.

\begin{figure*}[ht!]
\includegraphics[trim={1cm 0cm 1cm 0},clip,width=1.0\textwidth]{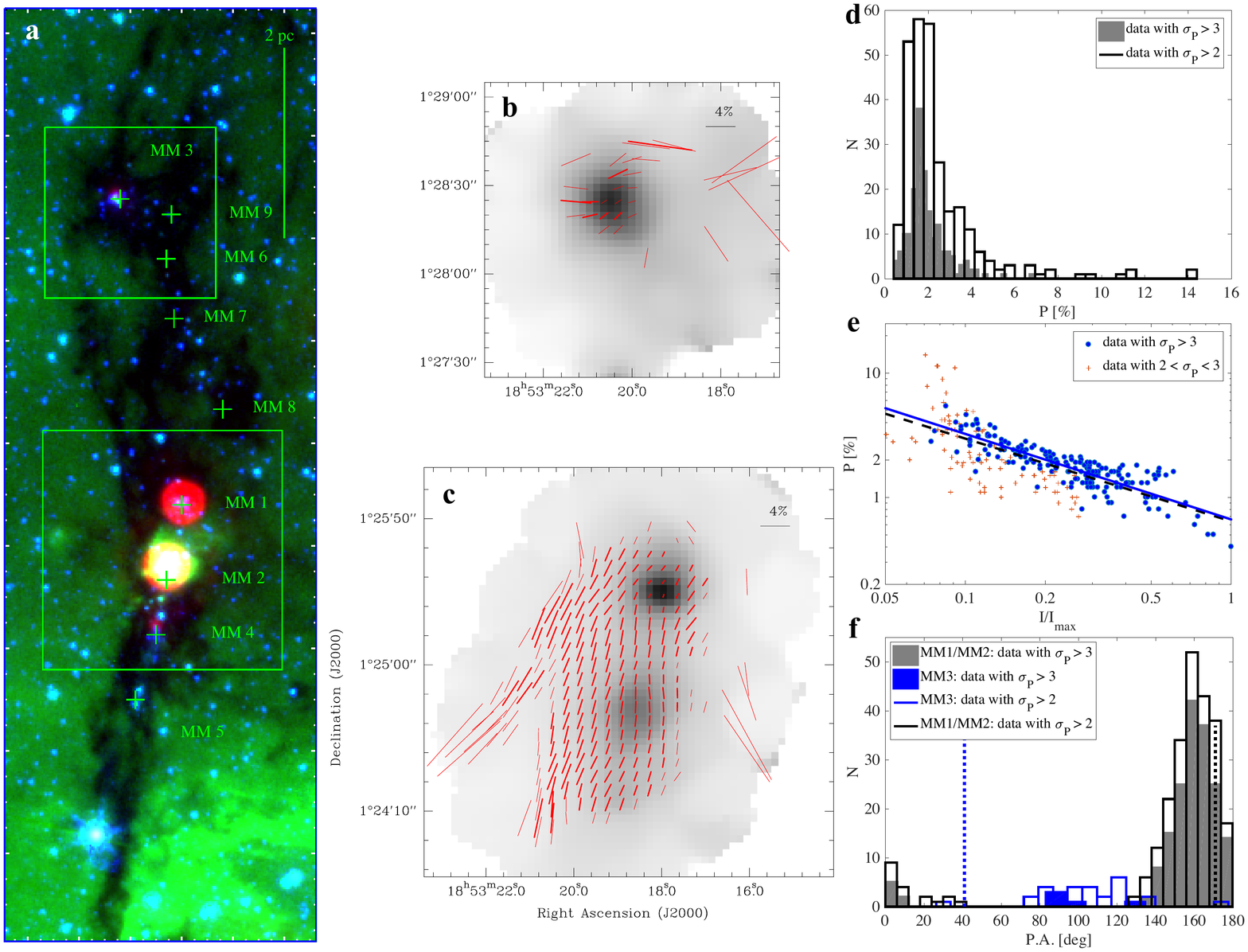}
\caption{a: G34 from {\it Spitzer} at 24 $\mu$m (red), 8 $\mu$m (green) and 4.5 $\mu$m (blue) (IRSA data archive). The green boxes mark the regions mapped with SHARP. The green pluses show the millimeter sources identified by \citet{2006Rathborne}.
b and c: Continuum emission at 350 $\mu$m (grey scale) and polarization (red segments) detected with SHARC and SHARP on the CSO toward MM1 and MM2 (panel c) and MM3 (panel b). The lengths of the polarization segments are scaled with the polarization percentages. 
d: Distribution of polarization percentages
for all cores, MM1, MM2, and MM3.
e: Stokes $I$, normalized to the peak emission of Stokes I ($I_{\rm max}$, the maximum in $I$ among all the three cores), versus polarization percentage $P$ for all the combined cores. 
Blue filled circles and red pluses display the data above $3\sigma_{\rm P}$ and between 2 to $3\sigma_{\rm P}$, respectively. 
The black solid and black dashed lines are power-law fits to the data above $3\sigma_{\rm P}$ and to the combined data (blue and red), yielding indices $-0.69$ and $-0.67$, respectively. f: Distributions of polarization position angles (P.A.), shown separately for the MM1/MM2 ridge and MM3. Note the clearly different prevailing orientations. The black dotted line and blue dotted line are the P.A. of the major axis of the MM1/2 of 171$\degr$ and MM3 of 41$\degr$, respectively, from the dendrogram analysis of the gas column density map derived from continuum emission (Peretto et al., in prep.).}\label{fig:pol}
\end{figure*}

In order to understand how stars form within filaments, studies of the environment and the structures of filaments in their early stages - typically as infrared dark clouds (IRDCs) - are essential.
%Within the well studied IRDCs, the G34.43 region appears particularly interesting due to the distributed low-mass stars along with massive stars at the center, suggesting that low-mass stars form earlier than, or contemporaneously with, high-mass stars in the G34 filament \citep{2014Foster}. 
The G34.43$+$00.24 filament (hereafter G34) displays an elongated and filamentary morphology with a length around 8 pc.
The distance of G34 is determined to be 3.59 kpc based on the detected velocity of N$_{2}$H$^{+}$ with respect to the standard of rest (Peretto et al., in prep.) and the Galactic rotation model \citep{2009Reid}. 
G34 harbors multiple cores, including G34-MM1 through MM9, that are likely at different evolutionary stages \citep{2011Chen}. 
Among these cores, MM2 has an associated UCHII region \citep{2004Shepherd}.
There are outflows detected originating from MM1, MM2, and MM3 \citep{2010Sanhueza,2008Rathborne,2013Sakai}, suggesting that these three cores are active star-forming sites.
The overall star-formation efficiency for the entire G34 filament is about 7\% \citep{2007Shepherd}. 
As suggested by the same authors, the MM1 - MM2 ridge may be magnetically subcritical so that the onset of core contraction and massive star formation is underway but was somewhat delayed. 
A recent study of the virial parameter $\alpha$, which quantifies the relative importance of the kinetic support against gravity, toward G34 indicates very small values, $\alpha\sim$0.2, across the filament, except at the southern and northern tips where $\alpha >$ 1\citep[][]{2014Foster}. 
This suggests that the filament is gravitationally bound.
Interestingly, two populations of stars are proposed to form in a sequence where low-mass stars are formed first, and then high-mass stars follow later \citep{2007Shepherd,2014Foster}. 
Combining the findings of low star-formation efficiency and small virial parameter, additional supporting mechanisms, such as the B field, are likely present to slow down the star-formation process in G34.

In order to study the impact of the B field on the G34 filament\footnote{Dust polarization at 850 $\mu$m with 14$\arcsec$ resolution of the G34 filament also has been observed with POL-2/JCMT during the revision of this manuscript. See Soam et al. (submitted) for details.} and its association with fragmentation, we observed the dust polarization at 350~$\mu$m using SHARP on the Caltech Submillimeter Observatory (CSO) in 2014 before its decommissioning.
We report the observations in section 2 and the results in section 3. 
The implications of these results together with a gas kinematics analysis in N$_2$H$^+$ and a comparison with higher angular resolution polarization images are discussed in section 4.
The conclusion is given in section 5.
\begin{figure*}
\includegraphics[trim={0 0.5cm 2cm 1cm},clip,scale=0.5]{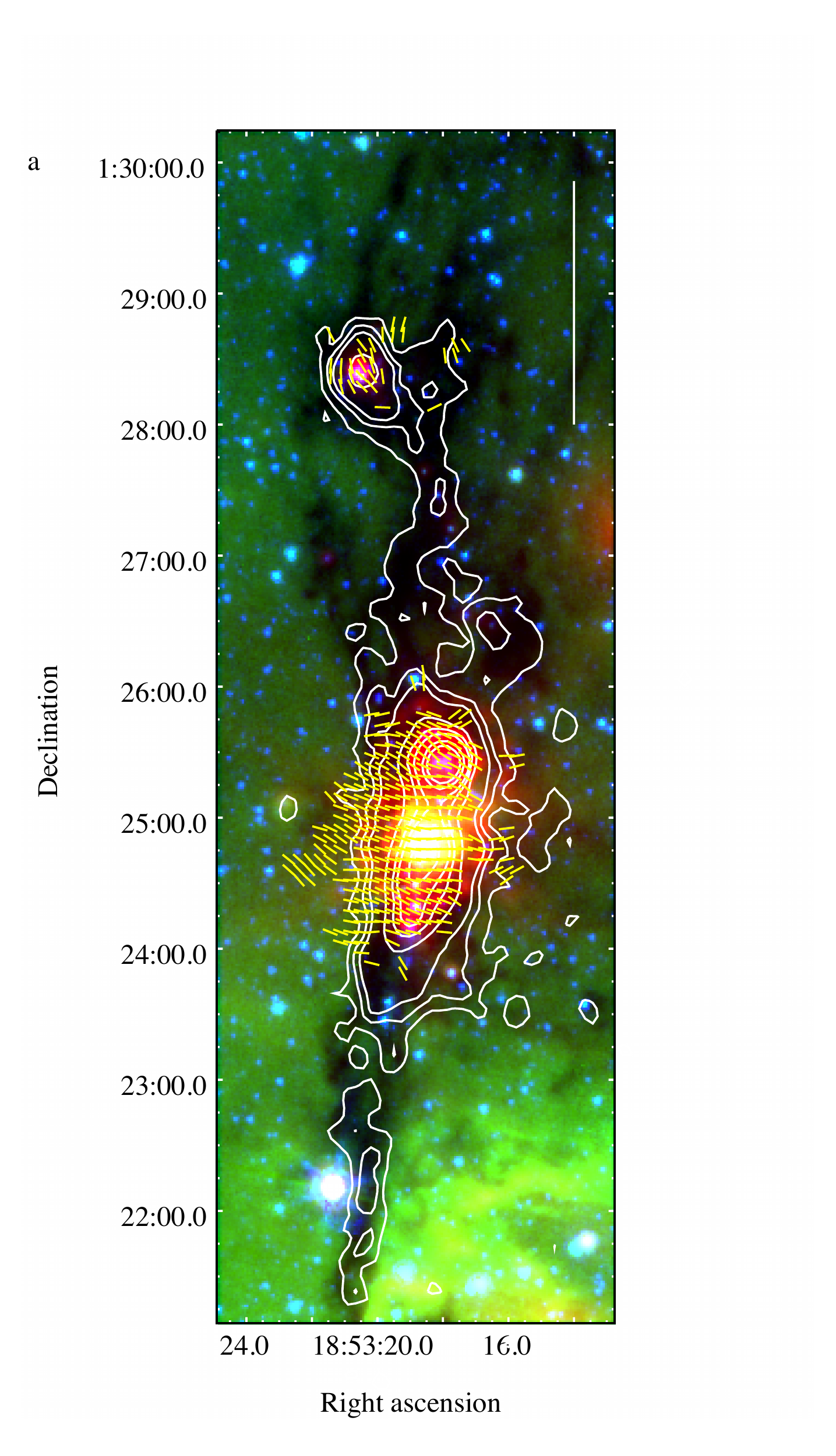}
\includegraphics[scale=.6]{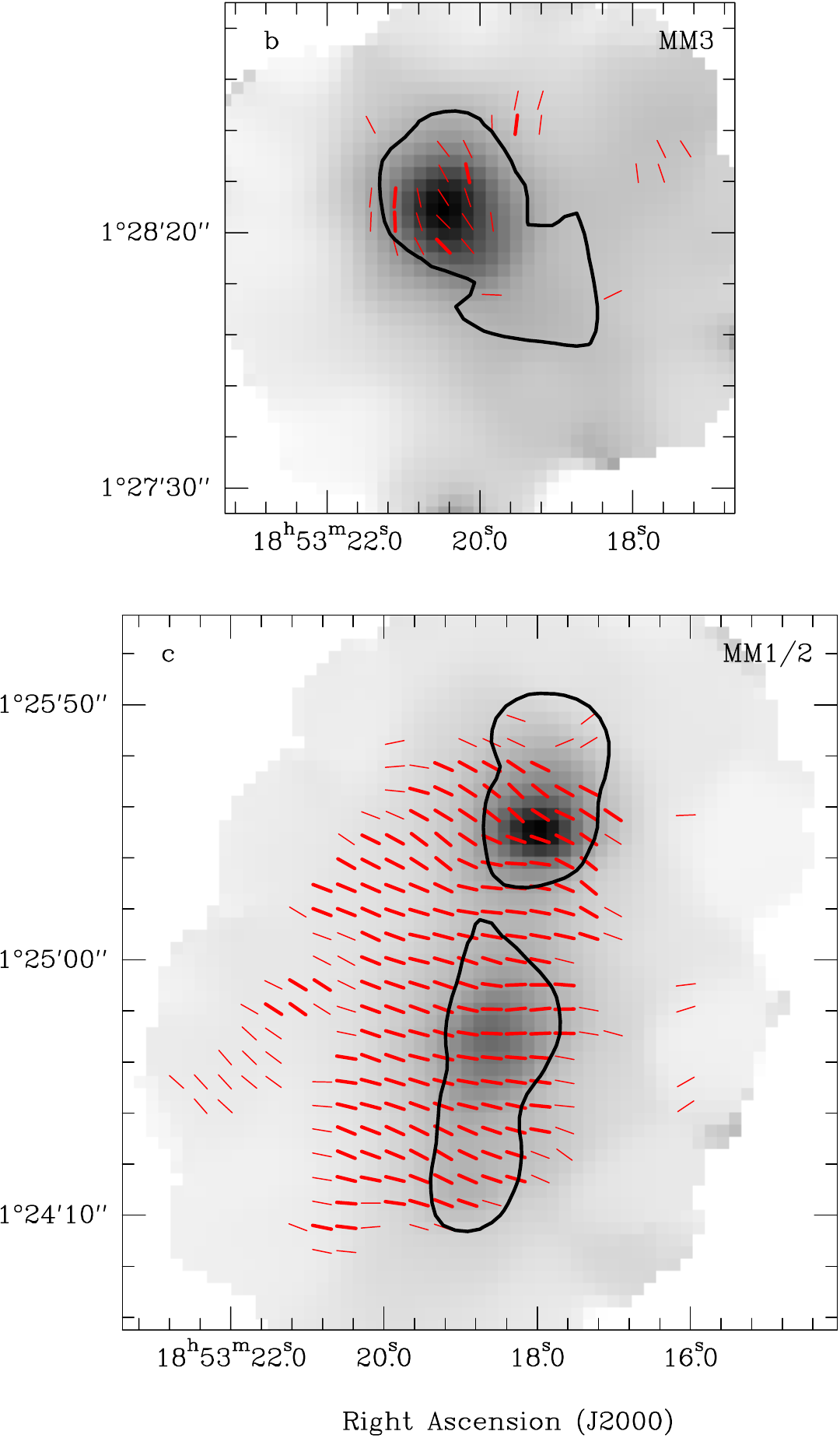}
\includegraphics[scale=.55]{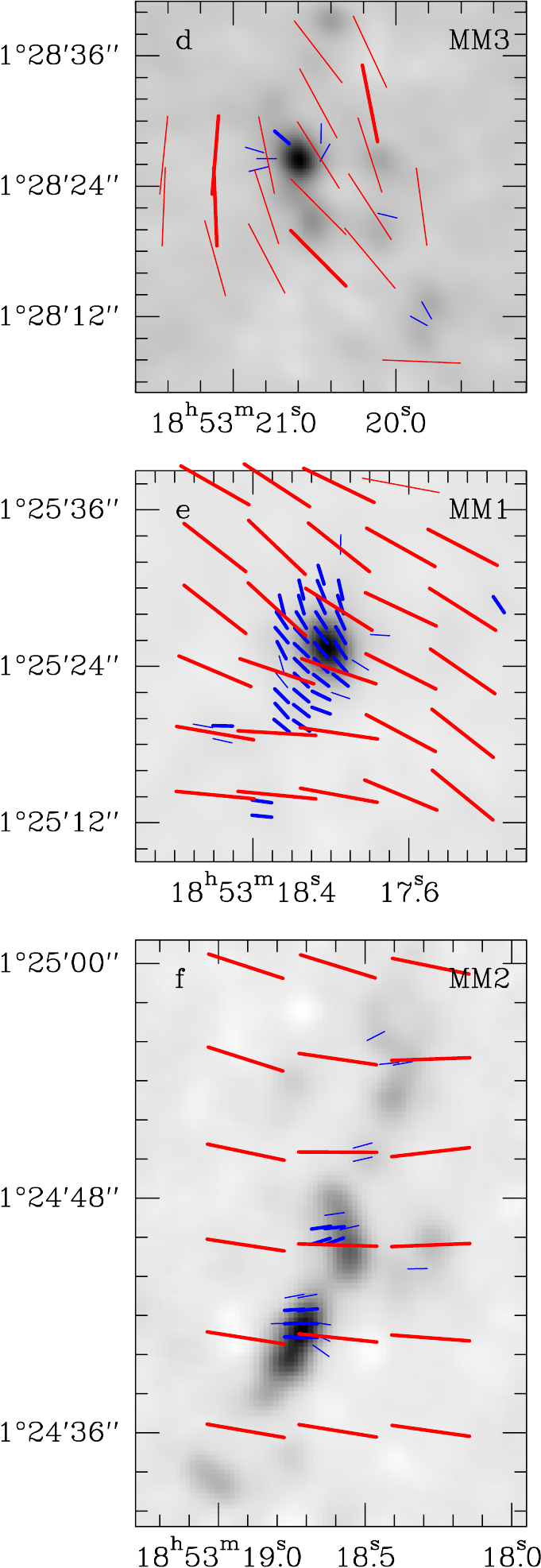}
\caption{B field map of the entire G34 filament from SHARP/CSO (overlaid on the 
Spitzer image as in figure 1a) in panel a, toward MM3 in panel b, toward MM1 and MM2 in panel c. Contours in panel a denote the 1.2 mm continuum emission at 60, 90, 120, 240, 360, 480, 840, 1200, and 2400 mJy beam$^{-1}$ \citep{2006Rathborne}. The contour in panel b and c marks the core region defined by the dendrogram analysis of the gas column density map (Peretto et al. 2019, in prep.).
Panels d-f display dust continuum (grey scale) and B field segments (blue segments) observed with higher angular resolution for MM3, MM1 \citep{2014Hull}, and MM2 \citep{2014Zhang}.
The red segments show the B field detected with SHARP, as shown in panel b and c.
In panels b-f, the dust continuum emission is shown in grey scale with dark color being strong emission.
}
\label{Fig:B_allscale} 
\end{figure*}
\section{Observations}
Polarimetric observations toward G34 were carried out using the SHARP polarimeter on the Caltech Submillimeter Observatory (CSO). 
SHARP \citep{2008Li} is a fore-optics module that adds polarimetric capabilities to SHARC-II, a 12 $\times$ 32 pixel bolometer array used on the CSO \citep{2003Dowell}. 
SHARP separates the incident radiation into two orthogonal polarization states that are then imaged side-by-side on the SHARC-II array. 
SHARP includes a half-wave plate (HWP) located upstream from the polarizing splitting optics. 
Polarimetric observations involve carrying out chop-nod photometry at each of the HWP rotation angles.
Each cycle requires approximately seven minutes. 
For a detailed description of the data analysis pipeline we refer to \citet{2013Chapman}.
Observations toward G34 were carried out at a wavelength of 350 $\mu$m with an effective beam size ($\theta$) of about 10$\arcsec$, on June 19, 20 and 22, and on July 3, 4 and 5, 2014, with a zenith atmosphere opacity at 225 GHz between 0.04 and 0.11.
The chop distance in cross-elevation was 300$\arcsec$.
We observed the G34 filament with four pointings, where three pointings covered the main dust ridge that surrounds MM1 and MM2, and a forth pointing was on the northern millimeter (mm) peak MM3. 
The pointing centers were at ($\alpha$, $\delta$) = (18:53:18.00, 01:25:25), (18:53:18.75, 01:24:57.5), (18:53:19.5, 01:24:30) for the MM1/MM2 ridge and (18:53:20, 01:28:15) for MM3.
The data were calibrated using the method described in \citet{2011Davidson} and \citet{2013Chapman}.
The polarization intensity shown in this paper has been debiased.  
The presented polarization results are gridded to half ($\sim5\arcsec$) of the effective SHARP beam resolution.
We note that no absolute flux calibration was routinely done during SHARP observations. The presented maps of the continuum emission are, thus,  all shown in relative strength.

The N$_{2}$H$^{+}$ J=1-0 observations of G34 were obtained as part of a larger IRAM-30m dense gas survey of IRDCs \citep[Peretto et al., in prep.,][]{2015Peretto}. 
At the frequency of the N$_{2}$H$^{+}$ J=1-0 transition, the IRAM-30m telescope provides a resolution of 27$\arcsec$. The $1\sigma$ noise (in Ta* scale) is ~0.1K per 0.16 km/s velocity channel and per 9$\arcsec$ pixel. The dense gas kinematics 
were extracted by fitting the hyperfine structures of the N$_{2}$H$^{+}$ line
%have been obtained by performing a fitting of the hyperfine structure of the line 
for every pixel using the GILDAS HFS line fitting routine. After visual inspection of the data cube, we concluded that only a single velocity component was needed to fit all positions in the cube. 
The results of this fitting are the N$_{2}$H$^{+}$ J=1-0 centroid and 
velocity dispersion maps for G34, which form the basis for the kinematic analysis presented in this paper.
These maps are shown in figure \ref{fig:n2hp}.
Spectra with a signal to noise ratio of less than 3 (estimated on the weakest of the HFS components) are masked out. 
As a result, the centroid velocity map we present here is very robust, with a typical velocity uncertainty $\simeq 0.1$ km s$^{-1}$.

%As a result of the fit, we have been able to compute the  N$_{2}$H$^{+}$ J=1-0 velocity centroid and velocity dispersion maps of G34, forming the basis for the kinematic analysis presented in this paper.

%
%
%
%
\section{Results}

\subsection{Dust Polarization at 350 $\mu$m wavelengths}\label{sec:result_pol}
The dust polarization intensity ($I_{\rm P}$) at 350 $\mu$m is clearly detected and resolved in the cores MM1, MM2, and MM3 of G34, in the fainter regions in between the cores, and also in the outer and more distant periphery toward the MM1 and MM2 ridge.
Polarization orientations are shown in figure \ref{fig:pol}b,c.
Thick segments denote a polarization signal larger than 3$\sigma_{\rm P}$, where $\sigma_{\rm P}$ is the noise of the polarization signal, and thin segments indicate data between 2 and 3 $\sigma_{\rm P}$. 
The median uncertainties in polarization orientations are 5.2$\degr$
and 10.9$\degr$ for the data $>3\sigma_{\rm P}$ and between 2 \textbf{and 3 $\sigma_{\rm P}$}, respectively.
Unweighted average and median polarization percentages over all cores are 
1.9\% (2.4\%) and 1.8\% (1.9\%)
with standard deviations, maximum and minimum values of 
0.9\% (2.0\%), 6.9\% (14.4\%) and 0.4\% (0.4\%) for data above 
$3~\sigma_{\rm P}$ and in parantheses when additionally taking into account data between $2~\sigma_{\rm P}$ and $3~\sigma_{\rm P}$.
Distributions of polarization percentage are approximately Gaussian centered around $\sim 2\%$ with a tail extending to $\sim 10-20\%$ (figure \ref{fig:pol}d).
%(figure \ref{fig:p_I_anti_correlation}).
%Polarization percentages vary from .... to ....\%, with a median value of ....\%. 
The largest polarization percentages $P$ 
are found in the faintest
Stokes $I$ regions, indicating that the generally observed $P-I$ anti-correlation also holds for filaments, here with a power-law exponent (fit without any weighting) around $-0.7$
%(figure \ref{fig:p_I_anti_correlation}).
(figure \ref{fig:pol}e).
Generally, polarization segments display only small variations in position angles (P.A.) between neighboring pixels. This is in particular the case around the MM1/MM2 ridge. 
Here, the average polarization orientation is 162$^{\circ}$ with a 
standard deviation of 14$^{\circ}$ (including all data above 
$2\sigma_{\rm P}$, figure \ref{fig:pol}f). 
This average orientation is calculated by redefining the few orientations between 0 and 40$^{\circ}$ to between 180$^{\circ}$ and 220$^{\circ}$. With this, a continuous
distribution without data across the $0^{\circ}/180^{\circ}$-ambiguity is achieved. 
The overall polarization orientation in MM3 is clearly oriented differently from MM1/MM2. Similarly redefining data yields an average orientation of 111$^{\circ}$ with a larger standard dispersion of $28^{\circ}$.

We note that calculating averages of circular (cyclic) quantities such as angles 
in the range between 0 and 360$^{\circ}$ or P.A. orientations between 0 and 180$^{\circ}$ is not well defined. Simple arithmetic averaging is generally not correct due to the cyclic nature. Redefining  
distributions as outlined above in order to avoid points of ambiguity is possible
for narrow or sufficiently confined distributions. 
One possible alternative approach is to convert all angles to their corresponding
locations on the unit circle and then calculate the average of these locations
\citep[for a complete reference on statistical analysis of circular data, see e.g.,][]{1995Fisher}. %https://en.wikipedia.org/wiki/Mean\textunderscore of\textunderscore circular\textunderscore quantities). 
In this approach, angles are interpreted as unit vectors. 
This approach is appropriate for broad distributions across points of 
ambiguity. 
%While this approach is 
%appropriate for broader distributions, its shortcoming is, e.g., that the mean 
%of 0$^{\circ}$ and 180$^{\circ}$ is not defined. 
For a comparison, we have also derived averages with this technique. In very good 
agreement with the above results we find average orientations of 157$^{\circ}$ and 106$^{\circ}$ for MM1/MM2 and MM3, respectively.

%An unambiguous average orientation is calculated by redefining the few orientations between 0 and 40$^{\circ}$ to between 180$^{\circ}$ and 220$^{\circ}$.
%The average polarization orientation (ignoring the few data points in the range between 0 and $50^{\circ}$, figure \ref{fig:pol}f), is 160$^{\circ}$ with a standard deviation of only $10^{\circ}$. 
%The overall polarization orientation in MM3 is clearly oriented differently from MM1/MM2 around 106$^{\circ}$ with a larger dispersion of $26^{\circ}$.

%around the MM1, MM2 and MM4 cores and also around MM3 (Figure \ref{fig:p_I_anti_correlation}) ({\bf ADD VALUES ?}).

The inferred B field map -- segments rotated by 90$^{\circ}$ with respect to the orientations of polarization -- is shown in figure \ref{Fig:B_allscale}. 
The long axis of the G34 filament is mostly along a north-south direction across the cores MM1 and MM2.
The B field across these cores is mostly perpendicular to the filament's axis, with a prevailing orientation around 70$^{\circ}$.
The MM3 core, oriented with its longer axis about 40$\degr$ off the north-south axis (figure \ref{fig:pol}f), reveals a B field that is changing its orientation to close to parallel to the core's dust 
ridge (figures \ref{Fig:B_allscale}a and b). 
These different B-field-versus-core orientations suggest different
roles of the B-field towards MM1/MM2 and MM3.

%Interestingly, the northern end of the filament near MM3 appears to be bent by at least 30$\degr$, with the B field changing its orientation to close to parallel to the longer axis of the MM3 dust ridge (figure \ref{fig:pol}f), 
%suggesting a different role of the B field toward MM3.

%
%
%
%
%
\begin{figure}
\includegraphics[width=0.45\textwidth]{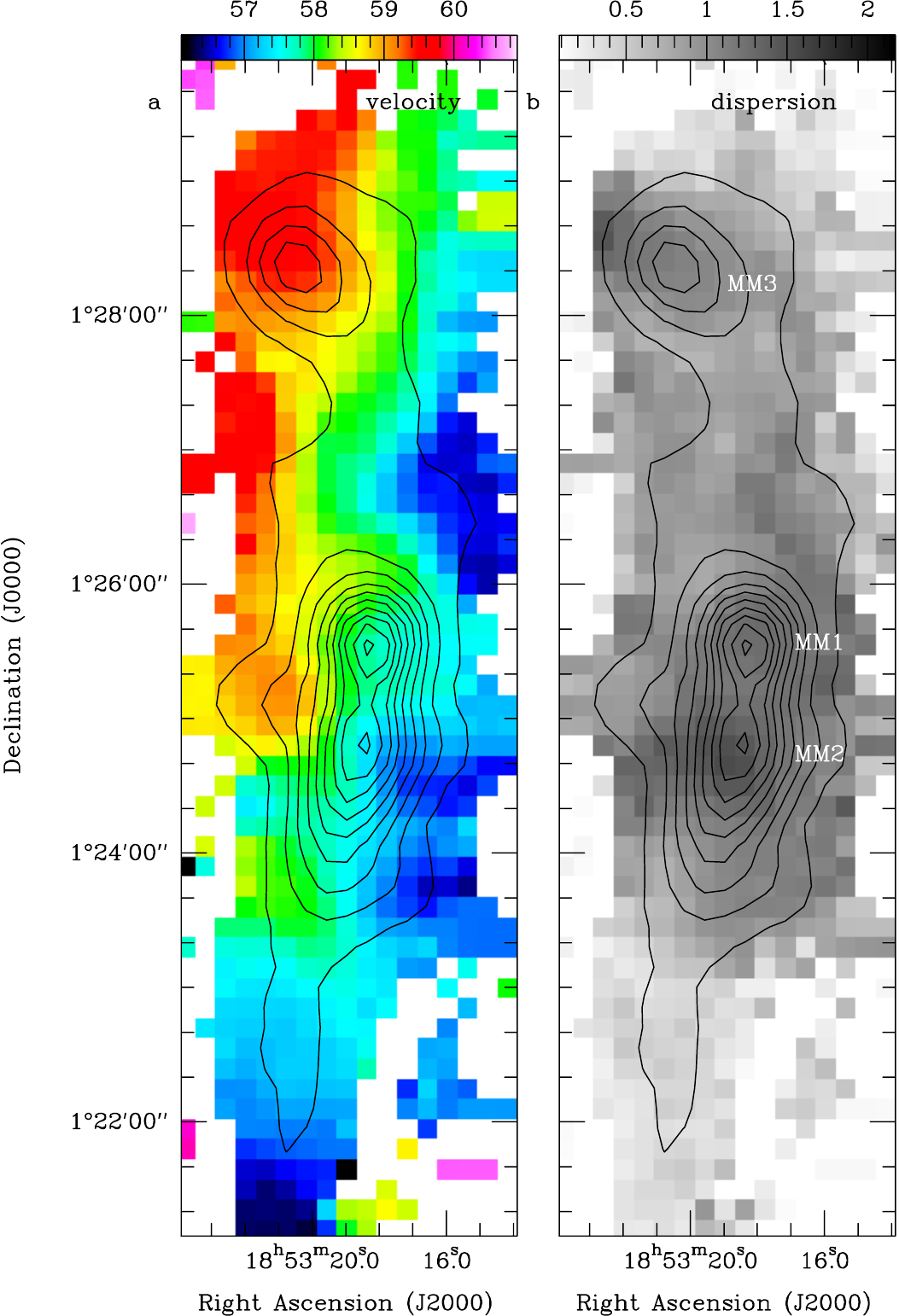}
\caption{The centroid velocity (a) and dispersion (b) of the detected N$_{2}$H$^{+}$ line in units of km s$^{-1}$ in color and grey scale. 
The contours denote the integrated emission of the N$_{2}$H$^{+}$ line.}\label{fig:n2hp} 
\end{figure}

At pc scale, the dust polarization in G34 at wavelengths around 3 mm was
observed with $\theta\sim17\arcsec$,
detecting 10 polarization segments
%at the wavelengths of 3 mm of G34 were observed with $\theta$ of $17\arcsec$, and 10 polarization segments were detected 
around MM1 and MM2 by \citet{2008Cortes}. 
At sub-pc scale, the B field toward MM1, MM2 and MM3 was resolved with $\theta\sim~$1$\arcsec$ with both the Sub-Millimeter Array (SMA) by \citet[][]{2014Zhang} and the Combined Array for Research in Millimeter-wave Astronomy (CARMA) by \citet[][]{2014Hull}. 
%This implies that the B field is well kept within the structure.
%The outflows of the dense cores are either parallel to perpendicular to the B field.
Comparing to the previous reported detections of dust polarization toward G34 by \citet{2008Cortes}, the results reported in this paper reveal the same overall B field orientations but a dramatically improved coverage across and along the filament.
\begin{figure}[tbh!]
\includegraphics[width=0.45\textwidth]{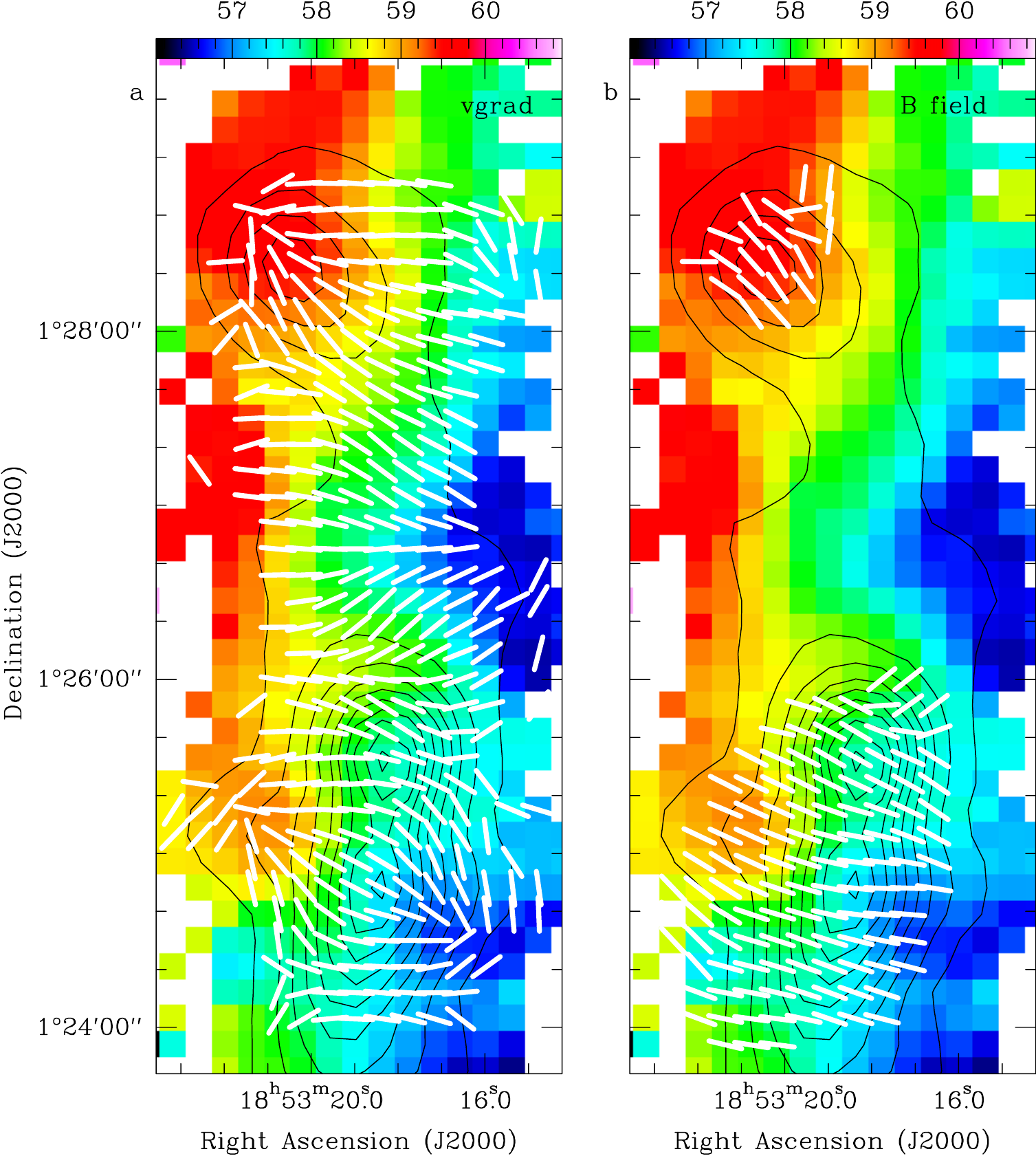} 
\caption{Maps of local velocity gradient orientations in white segments (panel a) and local B field orientations,  averaged over 27$\arcsec$ resolution, in white segments (panel b).
%Segments denote the angle of the velocity gradient in cyan (left panel) and of the B-field in red (right panel). 
Gas velocity is shown in color scale in both panels. 
Contours denote the integrated emission of the N$_{2}$H$^{+}$ line.
%The corresponding integrated intensity is in light-grey contours. 
%Left-hand-side of panel (a): local orientations of velocity gradients in black segments; right-hand-side: local B-field orientations (this paper). 
}
\label{Fig:bvcorr}
\end{figure}
\begin{figure}[tbh!]
\includegraphics[width=0.5\textwidth]{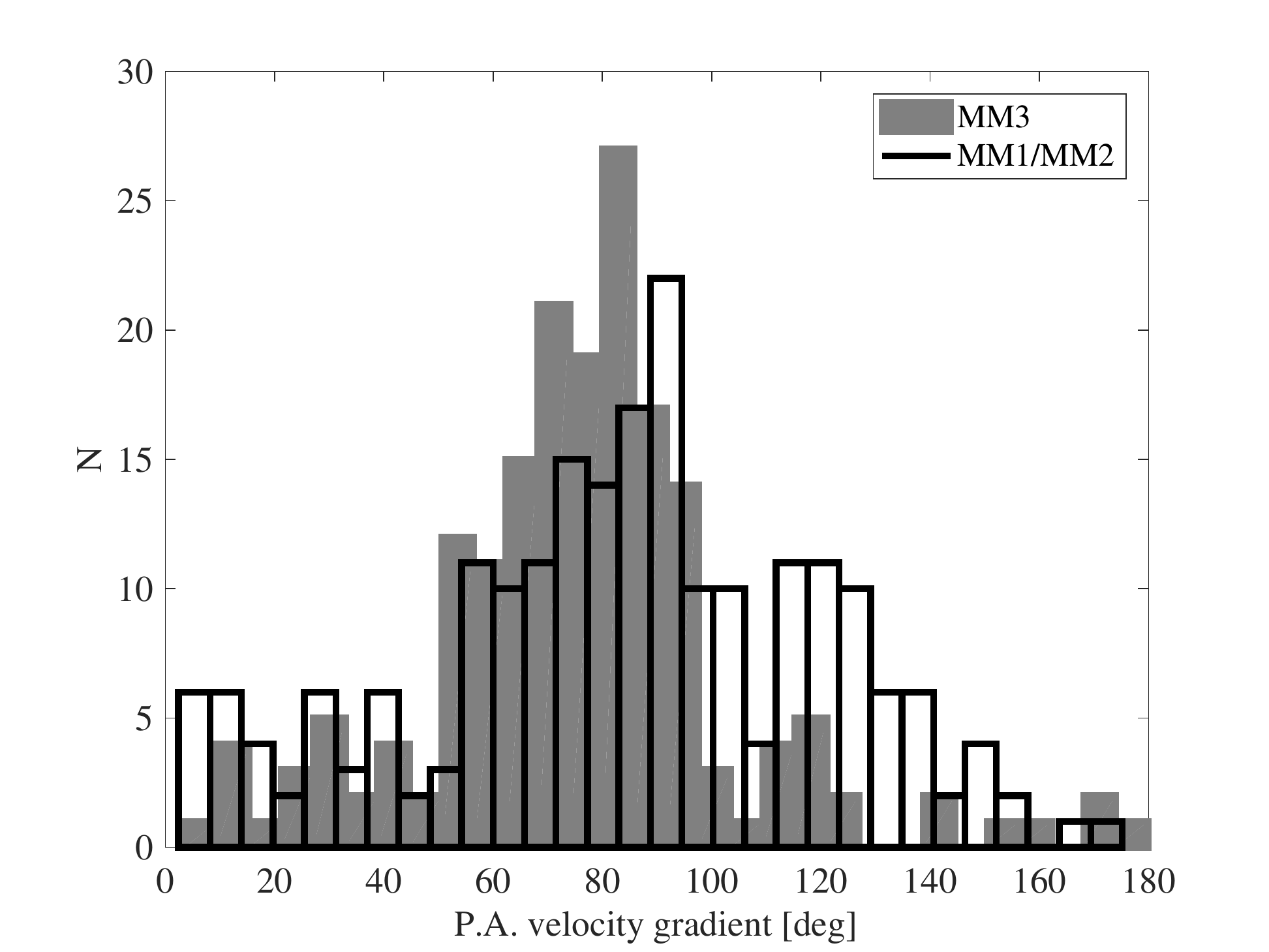} 
\includegraphics[width=0.5\textwidth]{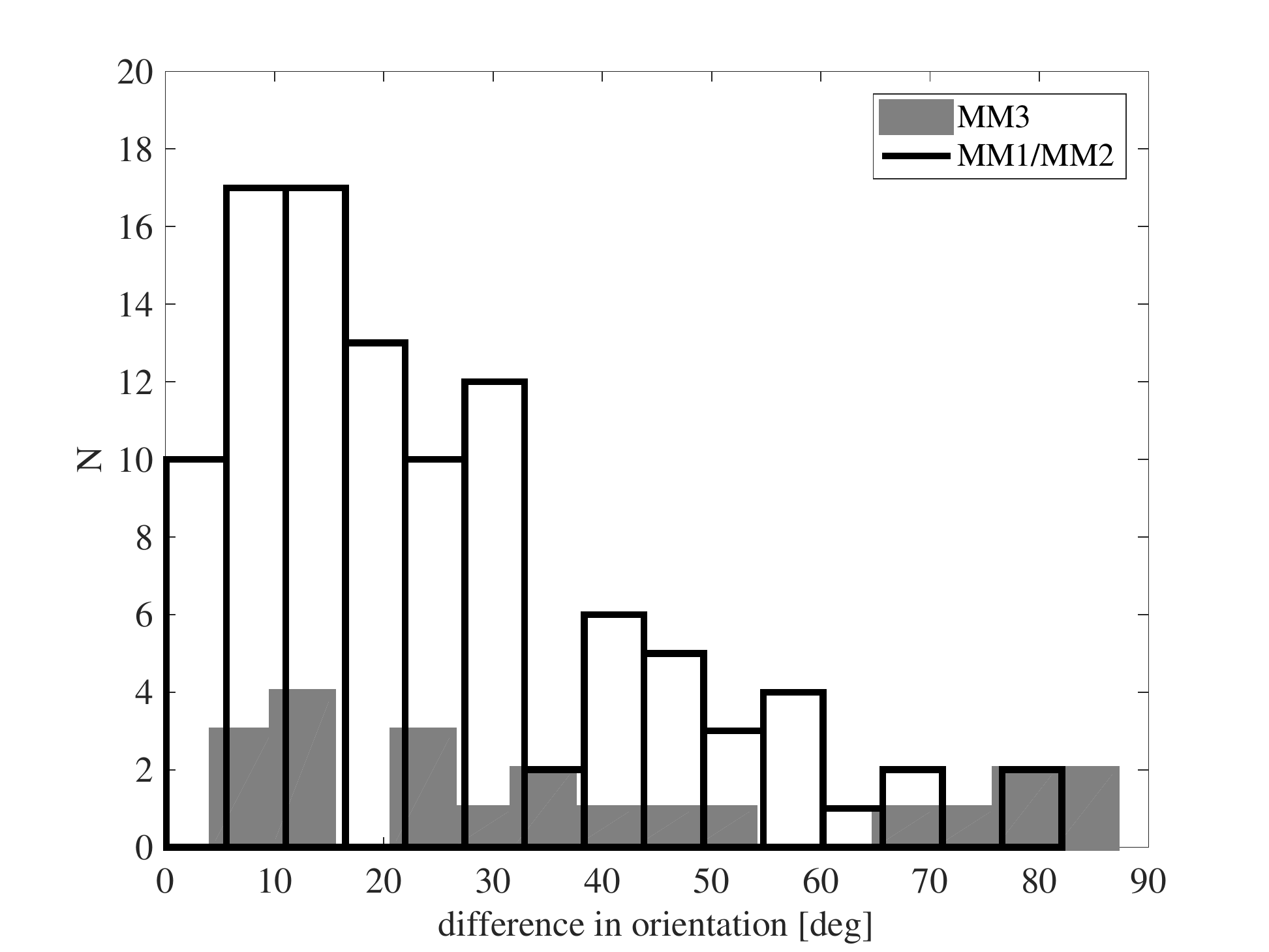} 
\caption{Upper panel: Histogram of the position angles of the velocity gradient orientations for data within the map sizes shown in panels b and c of figure \ref{Fig:B_allscale}. 
Lower panel: Histogram of P.A. differences between local
B field orientations, 
averaged over a 27\arcsec resolution,  and local velocity gradient orientations, where the two spatially overlap (panels (a) and (b) in Fig. \ref{Fig:bvcorr}).
Due to the resulting lower noise levels after averaging, all these data are $> 3\sigma_{\rm P}$.
The histograms shows that these angles are mostly well correlated, with differences in orientations below 20$^{\circ}$ for more than 50\% of the data.}
\label{Fig:plot_bvcorr}
\end{figure}
\begin{figure}[tbh!]
\includegraphics[width=0.3\textwidth]{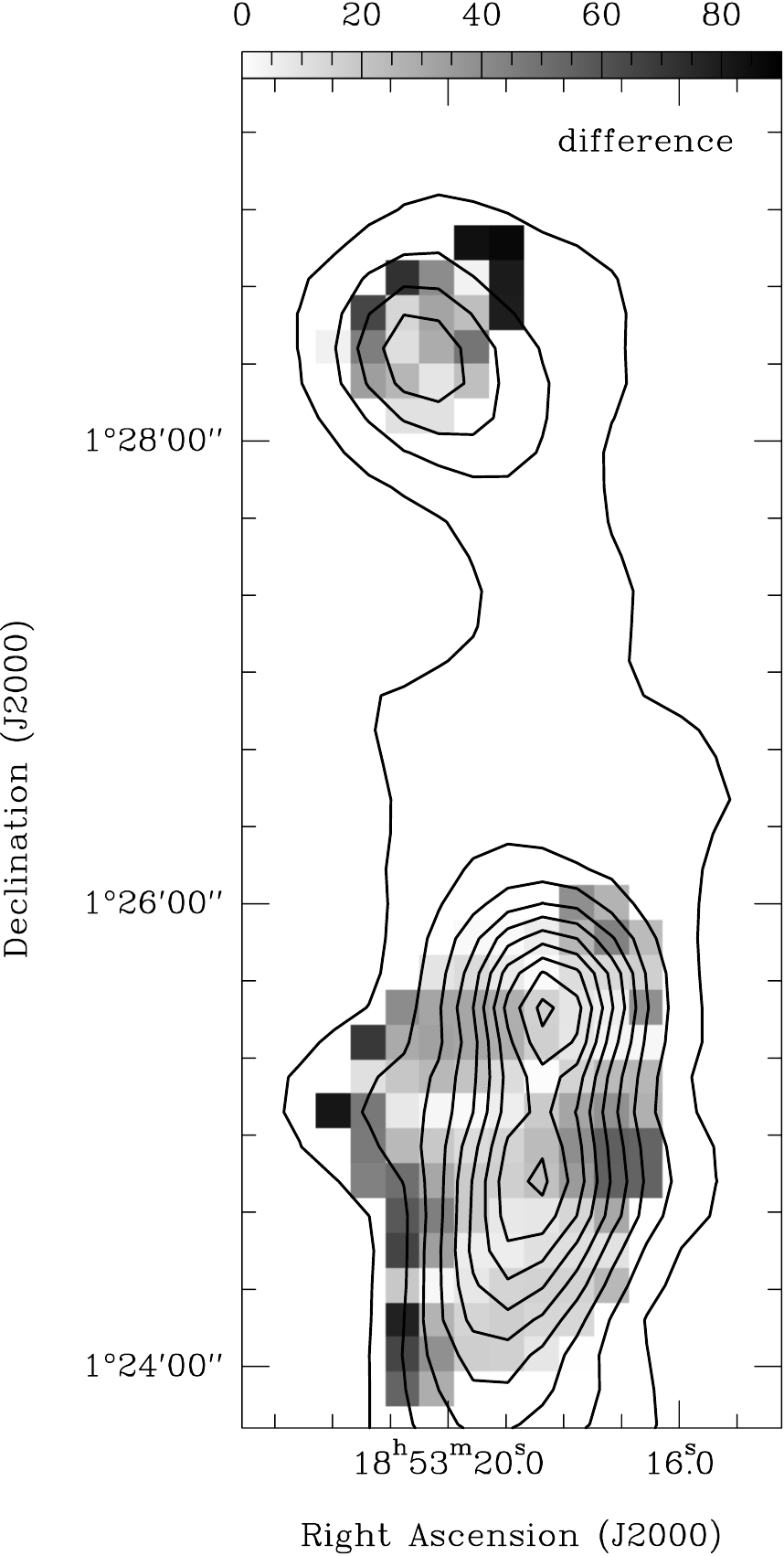} 
\includegraphics[width=0.5\textwidth]{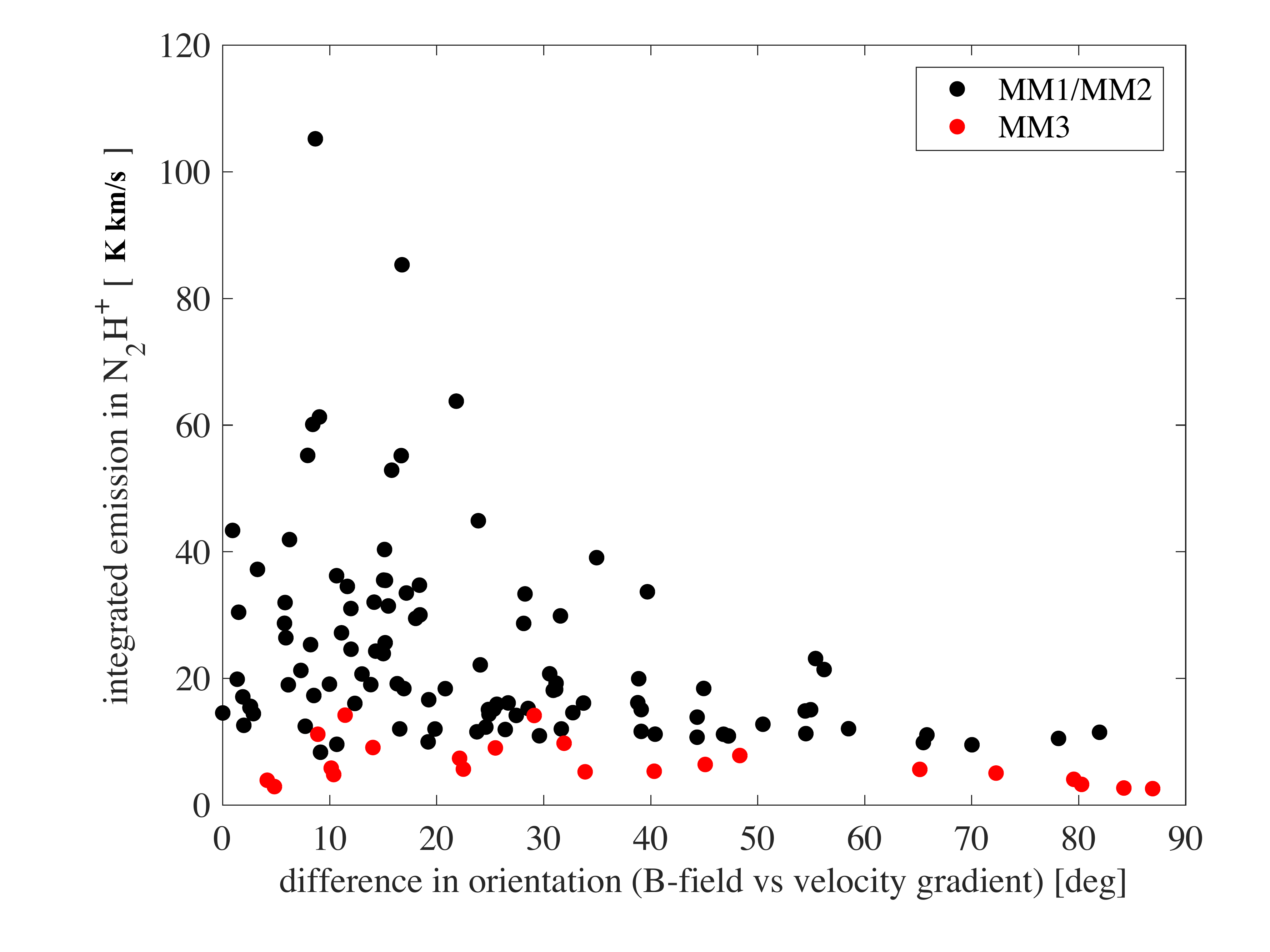}
\caption{Upper panel: Map of the angle difference between local B-field and velocity gradient orientations. Contours are the same as in figure \ref{fig:n2hp}.
Lower panel: Angle differences as a function of integrated N$_{2}$H$^{+}$ emission.
A clear trend is visible. With growing emission the angle differences are decreasing, likely indicating that an increasingly dominating gravity is aligning B field and 
velocity gradient. 
}
\label{Fig:map_bvdiff}
\end{figure}

\begin{figure}[tbh!]
\includegraphics[width=0.5\textwidth]{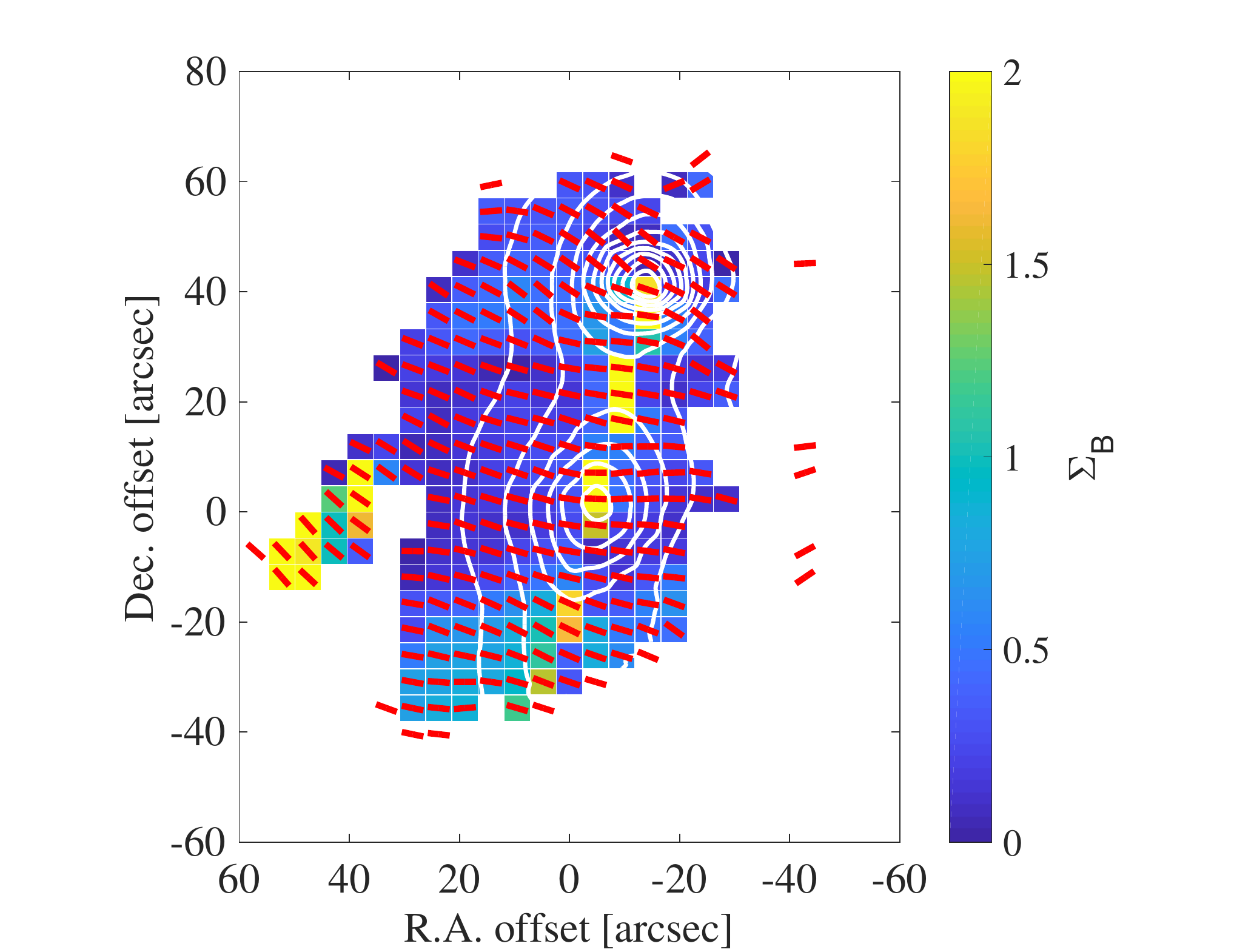} 
\caption{Spatial distribution of the B-field-to-gravity force ratio $\Sigma_{\rm B}$ using the method developed in \citet{2012Koch_a} toward MM1 and MM2. 
The offset is with respect to ($\alpha$,$\delta$)=(18:53:18.75, 01:24:57.5).
Contours denote the continuum emission at 350 $\mu$m. 
Red segments are the B field detected with SHARP with polarization signal $> 2\sigma_{\rm P}$. 
}
\label{Fig:sigmaB}
\end{figure}

\subsection{Connecting to Kinematics}
\label{sec:kinematics}
In order to understand the impact of the B field on the filament formation and to interpret the B field, we additionally extract kinematic information using the N$_{2}$H$^{+}$ (1-0) line.
The maps of the centroid velocity and the dispersion of the N$_{2}$H$^{+}$ line are shown in figure \ref{fig:n2hp}.
In terms of kinematics, G34 presents also a very organized velocity field throughout the 10 pc filament.
Overall, the velocity gradient is along an EW direction, consistent with measurements in NH$_{3}$ at a higher angular resolution of 3$\arcsec$ by \citet{2015Dirienzo}. 
The local velocity dispersion measured from N$_{2}$H$^{+}$ is largest near the UCHII region (i.e. MM2) and is smallest near MM3 (figure \ref{fig:n2hp}b).

We note that the measured velocity is along the line of sight.
In order to understand the gas motion across the plane of sky, we computed the local gradients of the line-of-sight velocity of G34.
The velocity gradient $(\partial v_z/\partial x, \partial v_z/\partial y)$ at each location is calculated by taking the differences of the velocities 2 pixels away from that pixel on each side in both right ascension and declination, where 1 pixel is 9$\arcsec$. 
We explicitly note that the local gradient is the change of the line-of-sight velocity $v_z$ (shown in color scale in figure \ref{fig:n2hp}a) in the plane of sky. 
It is, thus, a quantity in the same plane as the local B-field orientation.
The results are shown as segments in figure \ref{Fig:bvcorr}a.
Histograms of position angles of local velocity gradients are shown in the upper panel of figure \ref{Fig:plot_bvcorr}, where the data within the map regions of figure \ref{Fig:B_allscale}b and \ref{Fig:B_allscale}c are extracted.
The distributions peak around 85$\degr$ and 80$\degr$ toward MM1/MM2 and MM3, respectively.
Note that the prevailing orientation of the B field (section \ref{sec:result_pol}) is around 70$\degr$ and 20$\degr$ toward MM1/MM2 and MM3, respectively. 
Hence, local velocity gradients and local B field appear to be closely aligned toward MM1/MM2 but deviate more from each other toward MM3.
For the below comparison section \ref{sec:BV_gradient} between the velocity gradient and the B field, which is at a higher resolution $\theta\sim10\arcsec$, the polarization orientations are arithmetically averaged
to match the $27\arcsec$ resolution of the N$_{2}$H$^{+}$ data.
This lower resolution B field map is shown in figure \ref{Fig:bvcorr}b.

\begin{figure}[tbh!]
\includegraphics[width=0.45\textwidth]{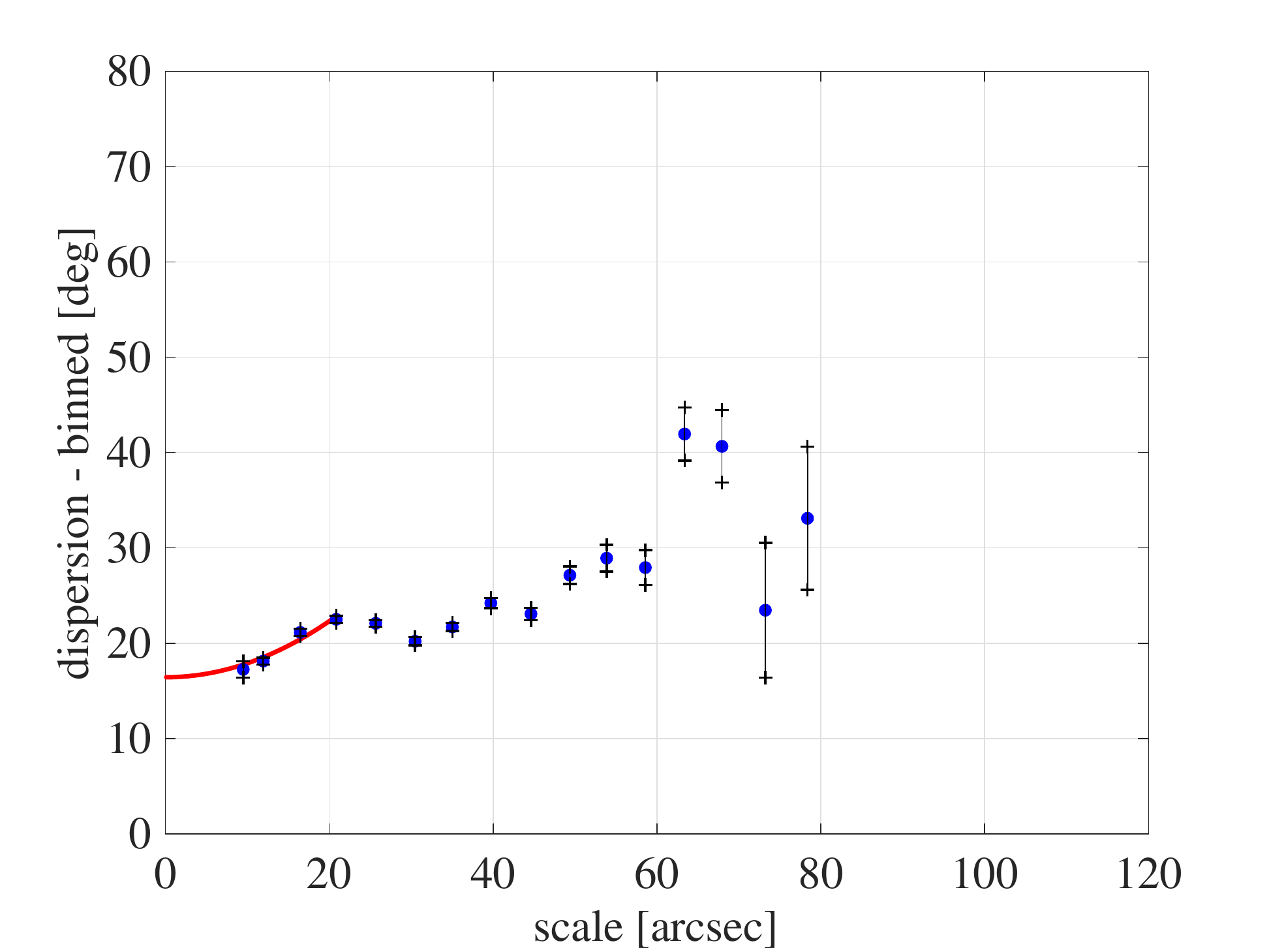}
\includegraphics[width=0.45\textwidth]{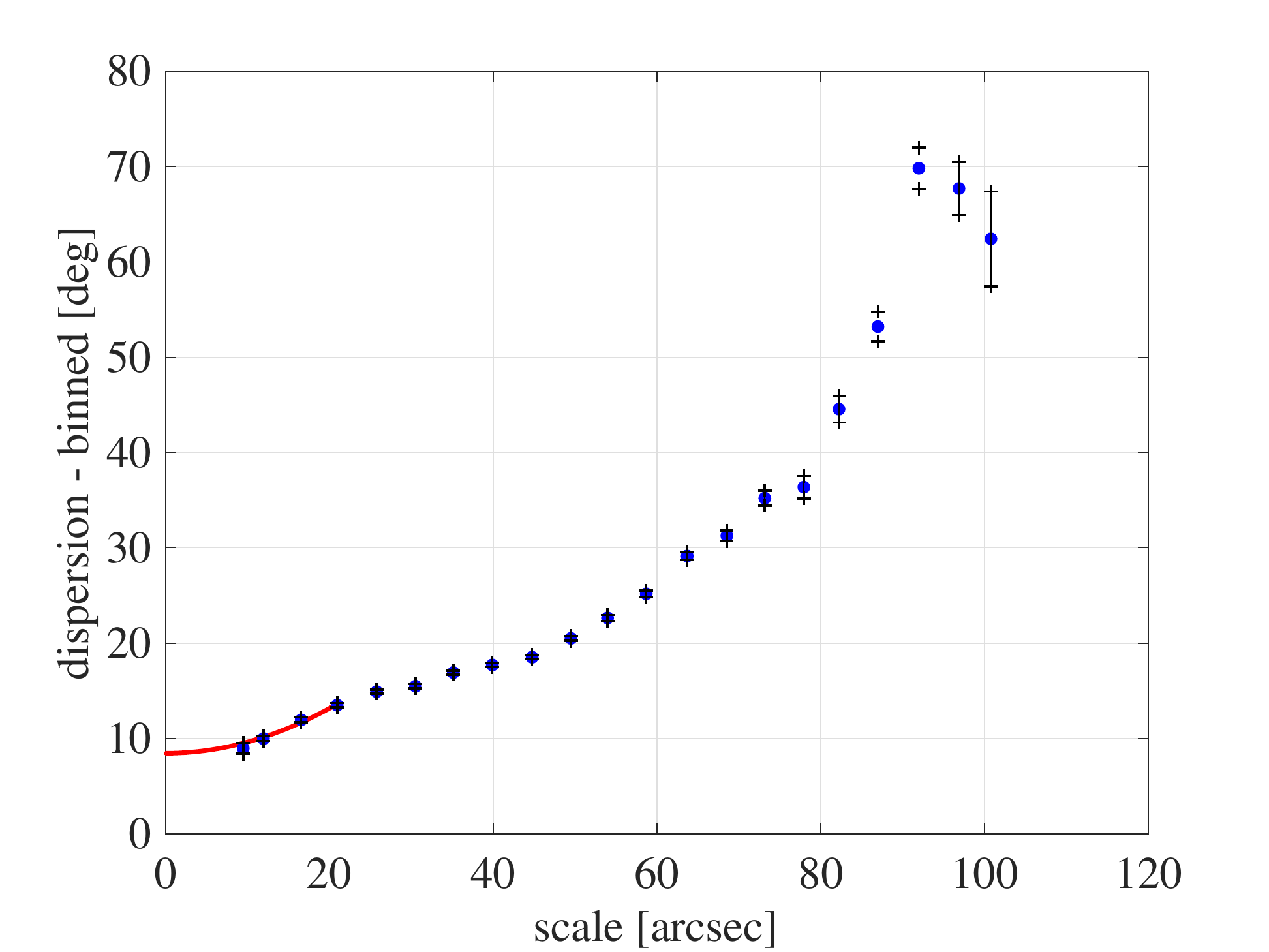}
\includegraphics[width=0.45\textwidth]{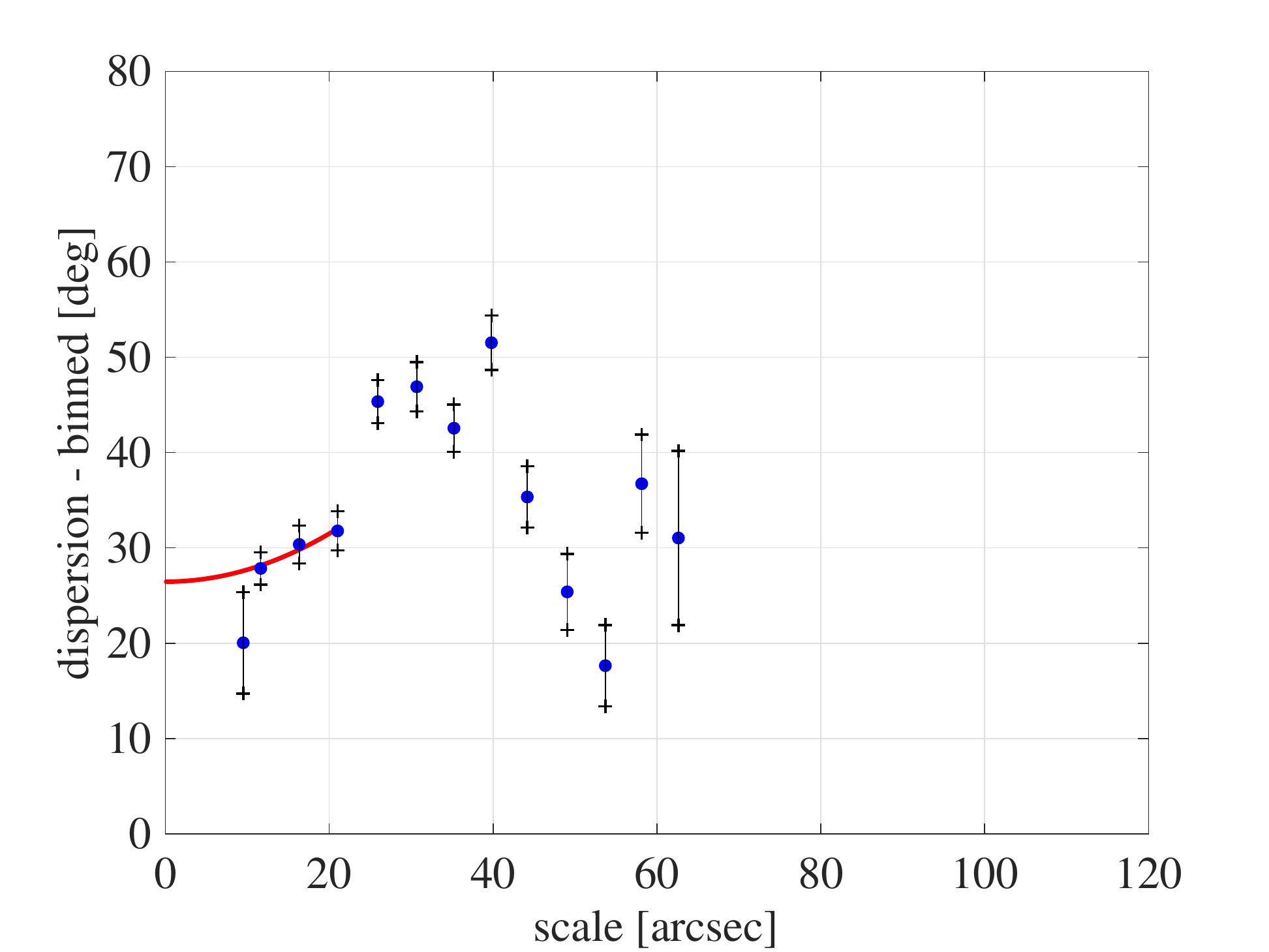}
\caption{Polarization dispersion functions from the larger clump areas of MM1, MM2, and MM3 from top to bottom, respectively. All polarization detections above $2\sigma_{\rm P}$ are included. 
The smallest scales are a measure of turbulence.
Quoted numbers for $\Delta\phi_{\rm B}$ in Table \ref{tab:cores} and \ref{tab:area} are the values at the smallest resolved scale around $10\arcsec$. %here $\phi_{\rm B}\approx 12\degr$.
The intercept (red fitted line at scale 0) quantifies the ratio of turbulent-to-mean magnetic field $\sqrt{\langle B_t^2\rangle}/B_0$ \citep{2009Hildebrand}.
Vertical bars indicate statistical uncertainties after binning and propagating the original measurement uncertainties. The small statistical uncertainties for MM1 and MM2 result from their small sample variances. 
The much smaller number of detections in MM3 leads to both a larger uncertainty and a much less regular structure function. }
\label{Fig:plot_sf}
\end{figure}

%
%
%
%section 4
%
\section{Discussion}
\subsection{B-Field--Velocity-Gradient Alignment}\label{sec:BV_gradient}
As shown in section 3, an organized B-field structure is detected along the entire filament and across all cores with $\theta\sim 10\arcsec$. 
The analysis of the N$_{2}$H$^{+}$ gas kinematics suggests that the gas might be moving across the filament in the East-West direction, based on the velocity gradient analysis (see section \ref{sec:kinematics}).

We further compare these local orientations of the velocity gradient vectors with the local orientations of the large-scale B field (figure \ref{Fig:bvcorr} and \ref{Fig:plot_bvcorr}).
We find that the local differences between these orientations are small. 
More than $50\%$ of the data have a difference below 20$^{\circ}$. 
The lower panel of figure \ref{Fig:plot_bvcorr} further illustrates this with a histogram of these differences 
with a mean (median) difference of $26^{\circ}$ ($20^{\circ}$)  over all three cores, 
and  $24^{\circ}$ ($19^{\circ}$) and  $38^{\circ}$ ($31^{\circ}$) for the MM1/MM2 region and MM3, respectively.
Thus, the local B field and the local velocity gradients are closely aligned.

Additionally, we display the spatial distribution of the differences of these orientations, covering MM1, MM2 and MM3 (figure \ref{Fig:map_bvdiff}). This map makes it evident that the -- already small -- differences are not randomly distributed. 
Larger angle differences between velocity and B-field are located mostly at the periphery of the filament.
This can occur when the gravitational pull is starting to dictate the direction of the gas flow,
but the B-field (while dragged along and being bent) is still resisting
alignment. 
Small differences are probably seen where gravity or external pressure is driving gas along field lines, or simply where gravity has
already aligned gas flow and magnetic field.
This first qualitative finding is suggesting that the angle difference might be
anti-correlated with the integrated intensity of N$_{2}$H$^{+}$, and hence with column density (figure \ref{Fig:map_bvdiff}, lower panel).
A quantitative comparison between B field, turbulence and gravity, which is discussed in the following subsections, is required to further test the proposed scenario of interpreting the angle differences.
In any case, our observations provide one of the first clear evidences of a direct local correlation between the magnetic field orientations and the observed velocity gradients at filamentary cloud scale.

%
%
%
%
%
% Table 1
\begin{deluxetable}{l|cccccc}[h!]
\tablecaption{Parameters of the core area -- 0.6 pc scale}
\tablehead{ \colhead{Object}  & \colhead{$M$}   &\colhead{$R$} & \colhead{$N_{\rm H_{2}}$} &  \colhead{$n_{\rm H_{2}}$} & \colhead{$\rho$} &  \colhead{$\triangle v$}\\
\colhead{}  &
\colhead{(M$_{\sun}$)}  &  
\colhead{(pc)} & 
\colhead{(${\rm cm}^{-2}$)} &  
\colhead{(${\rm cm^{-3}}$)} & 
\colhead{(${\rm g~cm^{-3}}$)}  &    \colhead{({\rm km s$^{-1}$})}} 
\startdata
%\multicolumn{13}{c}{2 pc scale, SHARP data}\\\hline
MM1 &   875 & 0.29 & 14$\cdot 10^{22}$ & 1.6$\cdot 10^{5}$  & 7.5$\cdot 10^{-19}$   &   1.1$\pm$0.1 \\
MM2 &   688 &  0.27 & 13$\cdot 10^{22}$& 1.6$\cdot 10^{5}$ & 7.5$\cdot 10^{-19}$ &   1.4$\pm$0.1\\
MM3 &   417 & 0.42 & 4$\cdot 10^{22}$&0.3$\cdot 10^{5}$ & 1.4$\cdot 10^{-19}$ &   0.9$\pm$0.1
\enddata
\label{tab:cores_def}
%\tablenotetext{Derived Properties of the cores}
\tablecomments{Columns are mass ($M$),  radius ($R$), averaged gas column density ($N_{\rm H_2}$), mean gas number density ($n_{\rm H_2}$), mean mass density ($\rho$) within $R$, and velocity dispersion
($\triangle v$).}
\end{deluxetable}
%

%
%
%
%
%
% Table 2
\begin{deluxetable}{l|cccccc}[h!]
\tablecaption{Parameters of the clump area -- 2 pc scale}
\tablehead{ \colhead{Object}  & \colhead{$M$}   &\colhead{$R$} & \colhead{$N_{\rm H_{2}}$} &  \colhead{$n_{\rm H_{2}}$} & \colhead{$\rho$} &  \colhead{$\triangle v$}\\
\colhead{}  &
\colhead{(M$_{\sun}$)}  &  
\colhead{(pc)} & 
\colhead{(${\rm cm}^{-2}$)} &  
\colhead{(${\rm cm^{-3}}$)} & 
\colhead{(${\rm g~cm^{-3}}$)}  &    \colhead{({\rm km s$^{-1}$})}} 
\startdata
%\multicolumn{13}{c}{1 pc scale, SHARP data}\\\hline
MM1 &   3184 & 0.86 & 6$\cdot 10^{22}$ & 2.3$\cdot 10^{4}$ &  1.1$\cdot 10^{-19}$ & 1.1$\pm$0.2\\
MM2 &   3794 &  0.98 & 6$\cdot 10^{22}$ & 1.9$\cdot 10^{4}$ & 0.9$\cdot 10^{-19}$ &  1.1$\pm$0.3\\
MM3 &   1224 & 0.98 & 2$\cdot 10^{22}$ & 0.6$\cdot 10^{4}$ & 0.3$\cdot 10^{-19}$ & 0.8$\pm$0.2
\enddata
\label{tab:areas_def}
%\tablenotetext{Derived Properties of the cores}
\tablecomments{The columns are the same as in table \ref{tab:cores_def}.}
\end{deluxetable}
%

%
%
%
%
%
% Table 3
\begin{deluxetable*}{l|ccccccccccccc}[h!]
\tablecaption{Parameters derived from the core area -- 0.6 pc scale}
\tablehead{ \colhead{Object}   & \colhead{$\frac{\sqrt{\langle B_t^2\rangle}}{B_0}$} & \colhead{$\frac{\sqrt{\langle B_t^2\rangle}}{B_0}_{\rm N}$} & \colhead{N} & \colhead{$\triangle\phi_{\rm B}$} & \colhead{$B_{\bot}$} & \colhead{$\langle\Sigma_{\rm B}\rangle$}  & \colhead{$\lambda_{\rm obs}$} & \colhead{$\alpha_{\rm vir}$}  & \colhead{$\alpha_{\rm B,vir}$} & \colhead{$P_{\rm T}$} & \colhead{$P_{\rm B}$} & \colhead{$u_{\rm G}$} & \colhead{Relative}\\
\colhead{}   & \colhead{} & \colhead{} &\colhead{} & \colhead{(\degr)} & \colhead{(mG)} & \colhead{} & \colhead{} & \colhead{} & \colhead{}&\multicolumn{3}{c}{(10$^{-9}$ dyn/cm$^2$)} & \colhead{importance}}
\startdata
%\multicolumn{13}{c}{1 pc scale, SHARP data}\\\hline
MM1  & 0.20$\pm$0.03  & 0.98$\pm0.22$ & 25$\pm13$ & 18$\pm$2 & 0.49$^{+0.11}_{-0.09}$ & 0.55$\pm0.04$   & 1.1$^{+0.3}_{-0.2}$ & 0.5$^{+0.1}_{-0.1}$ & 1.1$^{+0.4}_{-0.3}$ & 13.6$^{+2.6}_{-2.4}$ & 14.3$^{+7.2}_{-4.8}$ &  45.1 & G$>$B$\sim$T\\\\
MM2  & 0.15$\pm$0.03  & 0.54$\pm0.17$  & 14$\pm10$ & 10$\pm$2 & 1.12$^{+0.37}_{-0.26}$ &  0.68$\pm$0.05  & 0.5$^{+0.1}_{-0.2}$ & 0.9$^{+0.1}_{-0.1}$ & 4.9$^{+3.3}_{-1.7}$ & 22.0$^{+4.3}_{-3.9}$ & 74.9$^{+57.6}_{-30.8}$ & 35.0 & B$\geq$G$>$T\\\\
MM3  & 0.32$\pm$0.03 & 1.06$\pm0.46$ & 13$\pm 11$ & 20$\pm$5 & 0.16$^{+0.07}_{-0.05}$ & 0.66$\pm$0.05  & 0.9$^{+0.4}_{-0.3}$ &  0.9$^{+0.3}_{-0.2}$ & 2.1$^{+1.4}_{-0.8}$ & 1.7$^{+0.4}_{-0.3}$ & 1.5$^{+1.7}_{-0.8}$ & 2.3 & G$>$T$\sim$B
\enddata
\label{tab:cores}
%\tablenotetext{Derived Properties of the cores}
\tablecomments{Columns are turbulent-to-mean magnetic field ratio ($\sqrt{\langle B_t^2\rangle}/B_0$), turbulent-to-mean magnetic field ratio corrected for the number of turbulent cells ($(\sqrt{\langle B_t^2\rangle}/B_0)_{\rm N}$), number of turbulent cells within the beam (N), dispersion of the polarization position angles at the resolution scale of $9\farcs5$ ($\triangle\phi_{\rm B}$), B field strength in the plane of sky derived from the CF method ($B_{\bot}$), magnetic field-to-gravity force ratio $\Sigma_{\rm B}$ based on the intensity gradient method averaged over the core area ($\langle\Sigma_{\rm B}\rangle$), ratio of the observed mass-to-flux ratio and the critical mass-to-flux ratio ($\lambda_{\rm obs}$), virial parameter ($\alpha_{\rm vir}$), virial parameter taking into account additional B field support ($\alpha_{\rm B, vir}$), turbulent pressure ($P_{\rm T}$), B field pressure ($P_{\rm B}$), gravitational energy density ($u_{\rm G}$), and the relative importance between gravity (G), B field (B), and turbulence (T). All the values are derived from the core area (marked as contours in figure \ref{Fig:B_allscale}b,c) defined in the dendrogram analysis.
All the values related to the line data are derived from data above 3$\sigma$.
Note that the polarization data above 2$\sigma_{\rm P}$ are all included for better statistics. Within the core areas of MM1 and MM2, most polarization data are above 3$\sigma_{\rm P}$.
Uncertainties in $\langle\Sigma_{\rm B}\rangle$ are small due to the sample variance factor when averaging over the ensemble. Hence, they are larger for the core area than for the clump area. In order to calculate uncertainties in $(\sqrt{\langle B_t^2\rangle}/B_0)_{\rm N}$ and $N$, the errors in $P_{\rm T}$ and $P_{\rm B}$ are symmetrized using the means of their absolute values. Uncertainties for $(\sqrt{\langle B_t^2\rangle}/B_0)_{\rm N}$ are larger than for $\sqrt{\langle B_t^2\rangle}/B_0$ because they depend on $N$ which is calibrated against $\sqrt{P_{\rm T}/P_{\rm B}}$.} 
\end{deluxetable*}
%

%
%
%
%
%
% Table 4
\begin{deluxetable*}{l|ccccccccccccc}[h!]
\tablecaption{Parameters derived from the clump area -- 2 pc scale}
\tablehead{ \colhead{Object}  & \colhead{$\frac{\sqrt{\langle B_t^2\rangle}}{B_0}$} & \colhead{$\frac{\sqrt{\langle B_t^2\rangle}}{B_0}_{\rm N}$} & \colhead{N} & \colhead{$\triangle\phi_{\rm B}$} & \colhead{$B_{\bot}$} & \colhead{$\langle\Sigma_{\rm B}\rangle$}  & \colhead{$\lambda_{\rm obs}$} & \colhead{$\alpha_{\rm vir}$} & \colhead{$\alpha_{\rm B,vir}$} & \colhead{$P_{\rm T}$} & \colhead{$P_{\rm B}$} & \colhead{$u_{\rm G}$} & \colhead{Relative}\\
\colhead{}  & \colhead{} & \colhead{} &\colhead{} & \colhead{(\degr)} & \colhead{(mG)} & \colhead{} & \colhead{}  & \colhead{} &\colhead{} & \multicolumn{3}{c}{(10$^{-9}$ dyn/cm$^2$)} & \colhead{importance}} 
\startdata
%\multicolumn{13}{c}{1 pc scale, SHARP data}\\\hline
MM1  & 0.21$\pm$0.02 & 0.59$\pm0.18$  & 9$\pm6$  & 17.3$\pm$0.8 & 0.19$^{+0.05}_{-0.04}$ &  0.42$\pm$0.02  & 1.2$^{+0.3}_{-0.2}$ & 0.4$^{+0.1}_{-0.2}$ & 0.9$^{+0.5}_{-0.3}$ & 2.0$^{+0.8}_{-0.7}$  & 5.7$^{+3.5}_{-2.1}$ & 7.7 & G$\sim$B$>$T \\\\
MM2  & 0.11$\pm$0.01  & 0.33$\pm0.12$  & 10$\pm8$  & 9.0$\pm$0.5 & 0.34$^{+0.10}_{-0.09}$ &  0.89$\pm0.02$   & 0.6$^{+0.3}_{-0.1}$ & 0.4$^{+0.2}_{-0.2}$ & 2.4$^{+1.6}_{-1.1}$ & 2.0$^{+1.0}_{-0.9}$& 18.4$^{+12.4}_{-8.5}$ & 6.5 & B$>$G$>$T\\\\
MM3  & 0.35$\pm$0.03 & 0.71$\pm0.38$  & 5$\pm5$  & 20$\pm$5 & 0.06$^{+0.04}_{-0.03}$  & 0.63$\pm$0.04  & 1.3$^{+1.0}_{-0.9}$ &  0.6$^{+0.2}_{-0.3}$ & 1.2$^{+1.4}_{-0.7}$ & 0.3$^{+0.2}_{-0.2}$ & 0.6$^{+1.0}_{-0.5}$ & 0.7 & G$\sim$B$\sim$T
\enddata
\label{tab:area}
%\tablenotetext{Derived Properties of the cores}
\tablecomments{Columns are identical to table \ref{tab:cores}.  Here, $\langle\Sigma_{\rm B}\rangle$ is the averaged value over the clump area.}
\end{deluxetable*}
%

%
%
%
%section 4.2
%
%
\subsection{Magnetic Field, Gravity, and Turbulence on Filament-to-Core Scale}\label{sec:GBT}

What is the interplay between turbulence, magnetic field, and gravity from the observed filamentary ($\sim$ 8~pc) scale down to the resolved scale of the MM1, MM2 and MM3 cores ($\sim$ 0.3 pc in radius)? Is any of the constituents dominant, negligible, or are they equally important? Can we assess their relative importance in shaping smaller cores and fragmentation on the next smaller scale ($\sim$ 0.02~pc; panels d, e, f in figure \ref{Fig:B_allscale})? 
%\textbf{In order to compare the properties toward the MM1, MM2 and MM3 regions, we compare the measurables derived both from the core area define from the dendrogram analysis and a larger area where the MM1/MM2 region is splitted into the MM1 and MM2 "large area" with a straight line at the declination of 1$\degr$25$\arcmin$10$\arcsec$.}

We aim at comparing and quantifying the main constituents with various methods in this section. 
As the derived parameters can be sensitive to the selected area (see the discussion in section \ref{sec:robustness}), we now consider two different representative scales, namely the smaller core area at a scale of 0.6 pc and the larger clump area at a scale of 2 pc.
The core areas of MM1, MM2, and MM3, marked with contours in figure \ref{Fig:B_allscale}b and c, are determined using the dendrogram analysis of the derived gas column density $N_{\rm H_2}$, where $N_{\rm H_2}$ is derived from the dust continuum emission. See  Peretto et al. 2019, (in prep.) for details.
The resulting basic parameters determined within these core areas are given in table \ref{tab:cores_def}.
The clump areas are defined as the map region of figure \ref{Fig:B_allscale}b for MM3, and the map region in figure \ref{Fig:B_allscale}c with a cut in declination at 1$\degr$25$\arcmin$10$\arcsec$ in order to separate the MM1 and MM2 clump areas.
For the mass estimates at this larger scale, we further apply a mask to the $N_{\rm H_2}$ map following the detection threshold of N$_2$H$^{+}$ shown in figure \ref{fig:n2hp}, so that the mass and velocity information are extracted from the same region within this threshold. 
Table \ref{tab:areas_def} lists the basic parameters determined from these clump areas.
Unless specifically mentioned, the parameters shown in table \ref{tab:cores_def} and table \ref{tab:areas_def} are used to derived the values in table \ref{tab:cores} and \ref{tab:area}.
For both scales, magnetic field orientations with a detection above
$2\sigma_{\rm P}$ are included in the analysis. 
The impact of including or discarding data between $2\sigma_{\rm P}$ and $3\sigma_{\rm P}$ is addressed in section \ref{sec:robustness}.

\subsubsection{Turbulent-to-Mean Magnetic Field: Structure Function Analysis}\label{sec:turb_B}
We calculate a polarization structure function of second order (i.e., dispersion function) to derive the relative level of turbulence, following the technique by \citet{2009Hildebrand}. By analyzing the trend in the scale-dependent polarization dispersion, rather than the dispersion around a single mean value, this approach has the advantage that it filters out contributions to the dispersion resulting from larger-scale systematic changes in the field orientation.
Separate structure functions are derived for the MM1, MM2, and 
MM3 regions based on the detected polarization orientations. We fit the dispersion 
$\langle\Delta\phi_{\rm B}^2(l)\rangle^{1/2}$ over the smallest scales $l$ where the measured dispersion is given by 
Equation (3) in \citet{2009Hildebrand} with $\langle\Delta\phi_{\rm B}^2(l)\rangle=b^2+m^2l^2+
\sigma_{\rm M}^2$, where $\sigma_{\rm M}$ are the measurement uncertainties, and $m$ and $b$ are fitting parameters. In particular, extrapolating to the smallest measured scales with this fitting function leads to the intercept
$b$ at the scale $l\equiv 0$ which provides a measure for the turbulent-to-mean field strength \citep[][]{2009Hildebrand,2009Houde}:
\begin{equation}
\frac{\sqrt{\langle B^2_t\rangle}}{B_0}=\frac{b}{\sqrt{2-b^2}}.
\end{equation}
We note that the above approach and the derived estimates are only valid for the magnetic field 
component projected on the plane of the sky. 
Figure \ref{Fig:plot_sf} displays the structure function analysis for our three regions at the clump area of 2 pc scale.
The column  $\sqrt{\langle B_t^2}\rangle/B_0$ in table \ref{tab:cores} and \ref{tab:area} lists the turbulence levels toward the MM1, MM2, and MM3 regions at the two different scales. 
At the clump area of 2 pc scale, for the MM1 and MM2 region we find a relative turbulence level of 0.21 and 0.11, respectively. 
The MM3 region appears slightly more turbulent, being 0.35.
Similar values are found toward the core areas of 0.6 pc scale.

%We estimate the relative level of turbulence following . 
%A polarization structure function of second order (i.e., dispersion function) is calculated separately for the MM1/MM2 and MM3 regions, based on the detected polarization orientations in figure \ref{fig:pol}.  Extrapolating this function to the smallest scales 
%yields a measure for the statistically averaged turbulent-to-mean magnetic field ratio, $\sqrt{\langle B_t^2}\rangle/B_0$ \citep[][]{2009Hildebrand,2009Houde}.
%Figure \ref{Fig:plot_sf} demonstrates the structure function analysis toward the combined MM1/MM2 region.
%The column  $\sqrt{\langle B_t^2}\rangle/B_0$ in table \ref{tab:parameters} lists the turbulence levels toward the MM1, MM2, and MM3 cores. 
%For the MM1/MM2 region we find a relative turbulence level of 0.21 and 0.11 for MM1 and MM2, respectively. 
%The MM3 region appears slightly more turbulent, being 0.35.

We note three possible shortcomings in the above derivation. 
Firstly, while the polarization coverage on MM1/MM2 is excellent and very complete, MM3 is more sparsely covered.
This might leave some bias for the MM3 result as a more complete coverage might amplify the turbulence level or reduce it to the level in MM1/MM2.
Secondly, the size of the binning intervals in figure \ref{Fig:plot_sf}
can lead to changes of a few degrees in dispersion and a few percent
in the ratio. We have adopted the smallest scale to be the resolution 
of the SHARP observations while subsequent binning intervals are half of that resolution, in order to have a finer binning for scales that fall in between
multiples of the resolution scale. 
The corresponding plotted scales are the averaged scales  
of the data points falling into each bin. Thirdly, the most important
uncertainty in the above derived turbulence levels results from the 
a priori unknown number $N$ of turbulent cells within the beam of 
an observation. The turbulence level scales with $\sqrt{N}$ \citep{2009Houde,2010Koch}, where $N$ grows with the beam size. 
We will further address and limit this uncertainty in section \ref{sec:pressure_terms}.

\begin{figure*}[tbh!]
\includegraphics[trim={1.5cm 2cm 2cm 3cm},clip,width=0.5\textwidth]{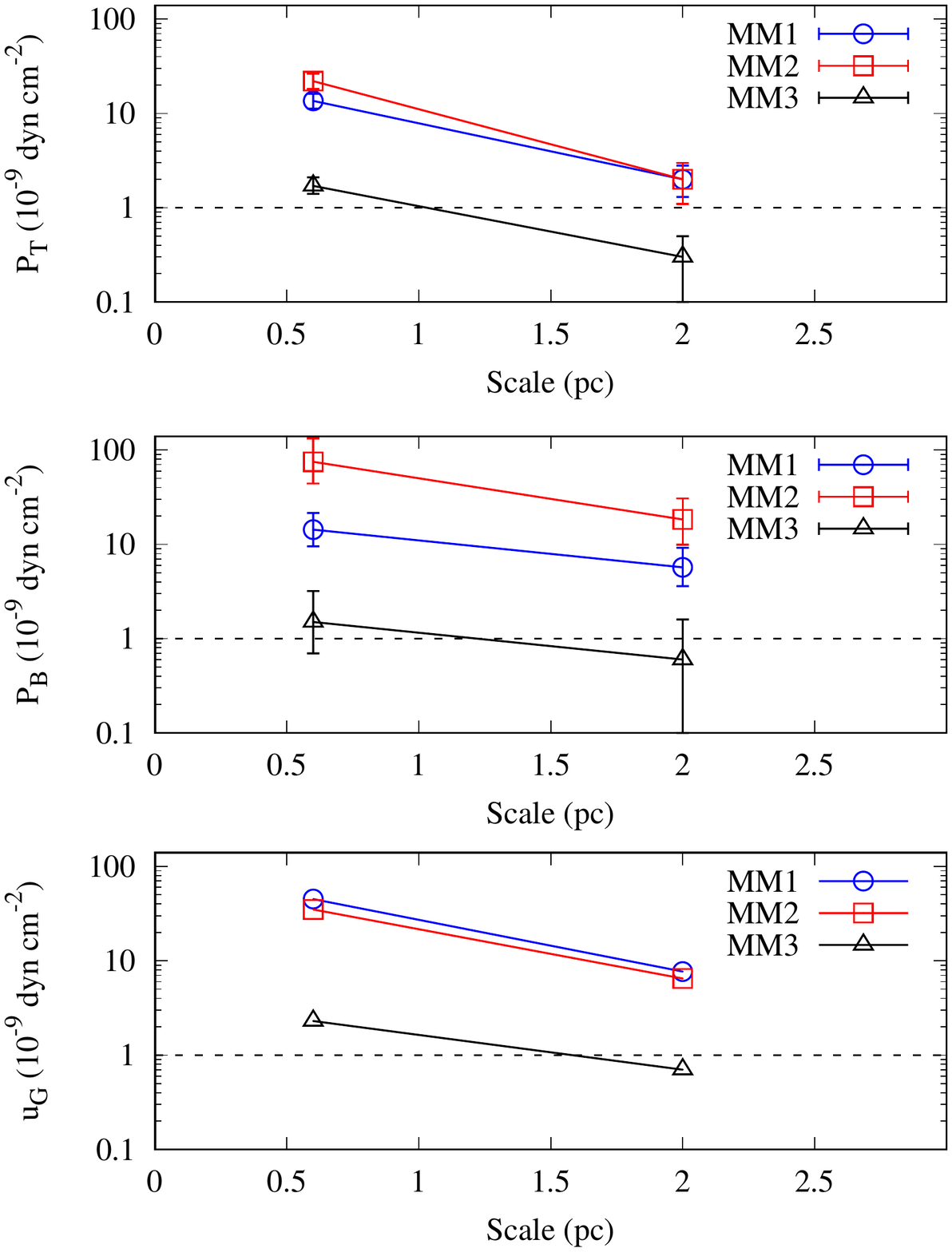}
\includegraphics[trim={1.5cm 2cm 2cm 3cm},clip,width=0.5\textwidth]{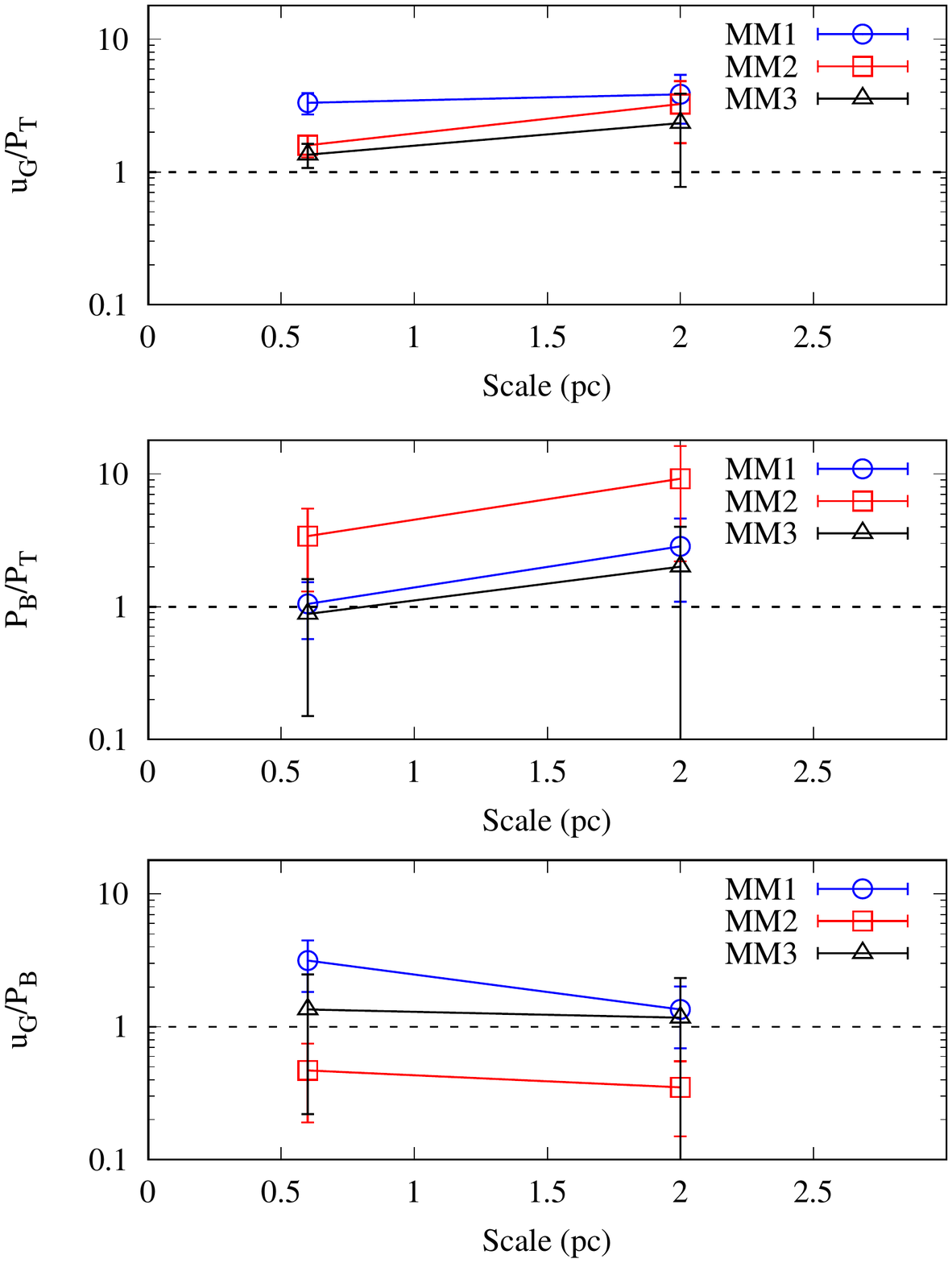}
\caption{Left panels: derived turbulent pressure ($P_{\rm T}$), magnetic field pressure ($P_{\rm B}$) and gravitational energy density ($u_{\rm G}$) for the larger clump scale (2 pc) and the smaller core scale (0.6 pc) for the three regions MM1, MM2, and MM3. 
Right panels: ratios among the three constituents:  $u_{\rm G}$/$P_{\rm T}$, $P_{\rm B}$/$P_{\rm T}$ and $u_{\rm G}/P_{\rm B}$. While a systematic and very similar growth for all three constituents is seen from clump to core scale in the left panels, the change in ratios over scale varies significantly for the three regions. 
}
\label{Fig:energ}
\end{figure*}

\subsubsection{Local B field-to-Gravity Force Ratio: Polarization-Intensity-Gradient Method}\label{sec:Sigma_B}
With the detected B field orientations and the associated dust continuum morphology, the polarization-intensity gradient method provides an estimate of the local field-to-gravity force ratio $\Sigma_{\rm B}$ 
\citep{2012Koch_a,2012Koch_b}.
In this technique, the MHD force equation is locally solved -- at every location where a polarization orientation is detected -- by identifying the various force terms with their corresponding directions on a map. The basic assumption is that an observed mass distribution (morphology) is the net result of the interaction 
of the various forces and that an intensity gradient can be associated with the direction of the inertial term in the MHD force equation. The validity of this assumption is further analyzed and demonstrated in \citet{2013Koch}.
The direction of the magnetic field tension force is orthogonal to an observed plane-of-sky B-field orientation. The direction of local gravity at a specific location is calculated by summing up all the surrounding pixelized emission. In the case of the MM1, MM2, and MM3 regions in G34, this emission is dust continuum (i.e. the total intensity of the continuum in Stokes I). In the polarization-intensity gradient method, only the local {\it direction} of gravity is needed but not its magnitude. It is, thus, sufficient to observe the dust morphology assuming that this is a fair approximation to the morphology (distribution) of the total mass. Thermal pressure is negligible in these cores\footnote{The temperature of the G34 filaments has been observed and analyzed in \citet{2015Dirienzo}, and the kinetic temperature of MM1 and MM2 is about the same, being 25 -28 K, while it is smaller toward MM3 (20 K). The contribution to the line broadening will be about 0.2 km s$^{-1}$, which is much smaller than the observed dispersion (being 0.8 km s$^{-1}$ or larger). The thermal pressure remains negligible. While on average thermal pressure is generally negligible, any significant local radiation pressure is not included in our analysis.}. 
The MHD force equation can then be solved geometrically by identifying two 
measurable angles between the above described orientations, namely
the angle $\delta$ between the gradient of the total intensity in Stokes I and the magnetic field 
orientation and the angle $\psi$ between the intensity gradient and the 
direction of local gravity. This
yields an expression for the field-to-gravity force ratio $\Sigma_{\rm B}$ \citep{2012Koch_a}:
\begin{equation}
    \Sigma_{\rm B}=\frac{\sin\psi}{\sin(\pi/2-\delta)}
                    =\frac{F_{\rm B}}{F_{\rm G}},
\end{equation}
where $F_{\rm B}=B^2/(4\pi R_{\rm B})$ and $F_{\rm G} = |\rho \nabla \phi|$
is the gravitational force with the potential $\phi$. When 
the density $\rho$ and the B field curvature $R_{\rm B}$ can also be determined, this technique leads to a map of B-field strengths. 
It needs to be emphasized that the force ratio $\Sigma_{\rm B}$ only relies on measurable angles. This is possible because the interplay of the various forces is encoded in the morphology (geometry and angles).
Moreover, $\Sigma_{\rm B}$ is minimally or not at all affected by projection effects as it is the ratio of two angles where uncertainties due to unknown line-of-sight inclinations can cancel out or be reduced \citep{2012Koch_a}.

Figure \ref{Fig:sigmaB} shows the $\Sigma_{\rm B}$-map for the MM1/MM2 region, and tables \ref{tab:cores} and \ref{tab:area} give the average values of $\Sigma_{\rm B}$ for the respective core and clump regions of MM1, MM2, and MM3. 
Similarly to the angle differences of the B-field and velocity orientations (figure 6), these force ratios are not random but appear in organized patterns with smooth changes. 
Gravity dominates over the B field from east and west in the peripheral zones, with the B field likely channeling material from the outside towards the filament
(local force ratio $\Sigma_{\rm B}<$1).
The B field resists gravity in between MM1 and MM2, and also at the southern ends and in the centers of the two cores, where gravity has not yet been able to pull in the field (local force ratio $\Sigma_{\rm B}>$1). The mean force ratio over the combined MM1/MM2 clump region is $\langle \Sigma_{\rm B}\rangle \sim$ 0.72. 
Highest values are seen in MM2, $\langle \Sigma_{\rm B}\rangle \sim$ 0.7 to 0.9 (see tables  \ref{tab:cores} and \ref{tab:area}), where also the largest
B field strength is measured (see section \ref{sec:DCF_B}). 
In conclusion, gravity dominates over the B field toward MM1 and MM2 on 
average,  while some specific locations show systematically larger values with $\Sigma_{\rm B}>$1.
While the B field is more aligned with the major axis in the MM3 core, gravity also dominates this core 
%but with a larger average field resistance, leading to a larger 
with a ratio 
$\langle \Sigma_{\rm B}\rangle \sim$ 0.6 on both clump and core scale.
We stress that although the average ratios are smaller than unity in all cores, 
there are clearly local zones with $\Sigma_{\rm B} > 1$ where the B field is dominating over gravity (see section \ref{sec:robustness} for a comparison with the mass-to-flux ratio). 

%
%
%
%
%
% section 4.2.3
\subsubsection{Magnetic Field Strength: Davis-Chandrasekhar-Fermi Method}\label{sec:DCF_B}

The absolute B field strength in the plane of sky, $B_{\bot}$, can be estimated from the Davis-Chandrasekhar-Fermi (DCF) method \citep{1951Davis,1953CF}. This technique is based on measuring and 
comparing turbulent gas motions with the resulting local dispersion in B field orientations. In this approach, large-scale components in both the velocity and the B field need to be removed in order to quantify the small-scale turbulent components. More recent numerical 
investigations suggest that the method is more reliably used if 
the dispersion in polarization angles is $<\sim$25$\degr$ \citep{2001Ostriker, 2008Diego}. Additionally, a numerical correction
factor on the order of unity is commonly added to the original DCF derivation 
to take into account inhomogeneities and line-of-sight averaging resulting from the three-dimensional B field-turbulence structures (e.g., \citet{2001Ostriker}). The DCF equation can then be written as
\citep{2004Crutcher}:

\begin{equation}
  B_{\bot} = A \sqrt{4\pi\rho} \frac{\triangle v}{\triangle \phi_{\rm B}},
\end{equation}
where $\rho$ is the mass density, $\triangle v$ is the turbulent velocity dispersion, and $\triangle\phi_{\rm B}$ is the dispersion in B field orientations. $A$ is the numerical correction factor and is adopted to be 0.5 \citep{2001Ostriker}.

%The absolute B field strengths in the plane of sky, B$_{\bot}$, can be estimated using the standard Chandrasekhar-Fermi method providing the dispersion in polarization angles being $<\sim$25$\degr$ \citep{1953CF, 2001Ostriker, 2008Diego}.
%We estimated the gas number density ($n_{\rm H_2}$) and the radius ($R$) for MM1, MM2, and MM3 from the dendrogram analysis on the Herschel data (Peretto et al., in prep.),  being {\bf 1.6$\times$10$^5$ cm$^{-3}$ and 0.29 pc for MM1, 1.6$\times$10$^5$ cm$^{-3}$ and 0.27 pc for MM1, and 3.0$\times$10$^4$ cm$^{-3}$ and 0.40 pc for MM3, respectively.
%$\rho$ is converted from the estimated $n_{\rm H_2}$}.
The gas velocity dispersion $\triangle v$ from Alv{\'e}nic gas motions is determined from the N$_{2}$H$^{+}$ map (figure \ref{fig:n2hp}b).
The uncertainties of $\triangle v$ only takes into account the dispersion of the values of $\triangle v$ within the selected areas, where the measurement uncertainties are not included.
%, being 1.2, 1.5 and 0.9 km/s when averaged over the cores of MM1, MM2, and MM3, respectively.
The dispersion in the B field position angles, $\triangle \phi_{\rm B}$, is estimated 
from the polarization structure functions (figure \ref{Fig:plot_sf}) 
adopting the smallest resolved scale around 10\arcsec. %$9\farcs5$.
%leading to 17$\degr$, 9$\degr$, and 20$\degr$ for MM1, MM2, and MM3, respectively  (Table \ref{tab:area}).
We note that working with the structure function, instead of calculating a single overall dispersion value in orientations, effectively separates large-
scale from small-scale changes in B field
orientations. This allows for a more accurate isolation of the
turbulent
dispersion without mixing changes in field orientations that are
driven by the large-scale B field morphology.
%The numerical factor $A$ 
The resulting uncertainties in the polarization dispersion are less than 1$\degr$ and about $2\degr$ for MM1 and MM2 on the clump and core scales, respectively, and they are about $5\degr$ for MM3 due to its fewer measurements. The uncertainties (including both $\triangle \phi_{\rm B}$ and $\triangle v$) are propagated to the resulting $B_{\bot}$ field strengths which are
0.19, 0.34, and 0.06 mG for the MM1, MM2, and MM3 clump areas, and 
0.49, 1.12, and 0.16 mG for their core areas (tables \ref{tab:cores} and \ref{tab:area}). 

%The resulting uncertainty of the dispersion of the polarization P.A. is less than 5$\degr$. 
%We thus neglect this uncertainty when we calculate the B field strength. 
%The resulting $B_{\bot}$ field strengths for the clump area in MM1, MM2, and MM3 are then 0.3, 0.6 and 0.2 mG, respectively.
%For the core area, the $B_{\bot}$ field strengths for MM1, MM2, and MM3 are 0.3, 0.7 and 0.1 mG, respectively.

%
%
%
%section 4.2.4
%
\subsubsection{Turbulent and B Field Pressure, and Gravitational Energy Density}\label{sec:pressure_terms}
The relative importance of turbulence, B field, and gravity can also be estimated from either their pressure terms or energy densities.
We estimate the pressure terms of B field and turbulence based on the 
quantities derived in the previous subsections.
The turbulent pressure can be calculated from 
\begin{equation}
P_{\rm T} = \frac{3}{2}\rho (\triangle v)^2, 
\end{equation}
where $\rho$ is the mass density and $\triangle v$ is the velocity dispersion along the line of sight. 
Here, the velocity dispersion is assumed to be isotropic, leading to the above factor $3/2$ for the total turbulent pressure.
The magnetic pressure term is calculated as
\begin{equation}
P_{\rm B} = \frac{1}{8} \pi B_{\rm total}^2. 
\end{equation}
%

%Assuming that the turbulent motion and the B field within the cores are isotropic, 
As the measured B field from polarization ($B_{\bot}$) is only the projected value in the plane of sky, the total B field strength ($B_{\rm total}$) is corrected for the statistical mean value, where $B_{\rm total}$ is 2$\cdot B_{\bot}$ 
\citep{2004Crutcher}.
%The turbulent pressure ($P_{\rm T}$) is assumed to be 3$\cdot P_{\rm T_\parallel}$, and the magnetic pressure ($P_{\rm B}$) is assumed to be 3/2$\cdot P_{\rm B_{\bot}}$. 
The gravitational energy density, $u_{\rm G}$ is calculated from 
\begin{equation}
u_{\rm G} = \frac{9}{20\pi} \cdot G \frac{M^2}{R^4} 
\end{equation}
assuming spherical geometry.
The derived $P_{\rm T}$, $P_{\rm B}$, and $u_{\rm G}$ are listed in the tables \ref{tab:cores} and  \ref{tab:area} for the clump and core scale.
Figure \ref{Fig:energ} shows the plots of the derived pressure terms at both scales and the ratios.
%\textbf{Interestingly, there is a clear difference in the role of gravity at these two scales. 
%At the core scale, the gravity appears to dominate over B field and turbulence. 
%At the clump scale, it appears that gravity is comparable to B field toward MM1. Toward MM3, gravity still dominates, while B field dominates over gravity toward MM2.} 

We can now compare the pressure ratio between turbulence and B field with the
ratio of turbulent-to-mean magnetic field strength
from the polarization structure function analysis in section \ref{sec:turb_B},
%-- which is solely based on the polarization orientations. 
where the latter one is solely based on the polarization orientations.
The apparent inconsistency between the numbers derived from the two
techniques can be explained by uncertainty in the number of turbulent cells $N$,
 because the turbulent-to-mean field ratio corrected for $N$ scales as 
$\sqrt{N/2}\cdot b$, where $b$ is the intercept of the structure function 
at scale 0 \citep{2009Houde,2010Koch}, derived as in figure \ref{Fig:plot_sf}. 
%We note that the B field strength derived from the DCF method involves a ratio between a turbulent velocity dispersion and a field dispersion such that identical beam integration effects will cancel out. 
Adopting the ratio
$P_{\rm T}/P_{\rm B}$, we can now estimate the number of turbulent cells
in our beam by scaling $\sqrt{\langle B_t^2\rangle^{1/2}/B_0}$ with $N$
to match $P_{\rm T}/P_{\rm B}$. 
This yields $N$ = 25, 14, and 13 for MM1, MM2, and MM3, respectively, 
for the core-scale data. The number of turbulent cells within the angular resolution is very similar for MM2 and MM3, and about twice larger in MM1.
$N$ for the clump scale appears to be smaller but consistent with the core-scale
estimates within uncertainties.
We note that when converting from pressure to field strength, a square-root operation is involved. 
%\textbf{An additional caution is that the unknown inclination of the actual B field geometry between MM1, MM2 and MM3 can vary the calculated $N$ values by a factor of a few.}
Values of turbulent-to-mean magnetic field strength ratios corrected for $N$ are also listed for both scales in table \ref{tab:cores} and \ref{tab:area}.

%\textbf{This yields $N$ = 21, 22, and 11 for MM1, MM2, and MM3, respectively.
%The number of turbulent cells within the angular resolution is very similar between MM1 and MM2, and about twice smaller in MM3.}
%We note that when converting from pressure to field strength, a square-root operation is involved. 
%Values of turbulent-to-mean magnetic field strength ratios corrected for $N$ are also listed in table \ref{tab:cores} and \ref{tab:area}.

%
%
%
% section 4.2.5
%
%
\subsubsection{Mass-to-Flux Ratio and Virial Parameter}
The relative importance of the magnetic field with respect to gravity can also be estimated by the mass-to-flux ratio with respect to the critical mass-to-flux ratio, 
\begin{equation}
    \lambda \equiv \frac{(M/\phi)_{\rm obs}}{(M/\phi)_{\rm critical}} = 7.6 \times 10^{-21}\frac{N_{\rm H_2}}{B_{\rm total}},
\end{equation}
 where $N_{\rm H_2}$ is the gas column density in cm$^{-2}$, and $B_{\rm total}$ is the total B field strength in $\mu$G \citep{2004Crutcher} as introduced in the previous sub-section, i.e. 
%Here, $N_{\rm H_2}$ \textbf{of the core area} is calculated by multiplying $n_{\rm H_2}$ with the radius $R$ for the line-of-sight extension.
$B_{\rm total}$ is estimated from the DCF method (see section \ref{sec:DCF_B}) with a correction for the statistical mean value ($B_{\rm total}=2\cdot B_{\rm \bot}$).
%The derived $\lambda_{\rm obs}$s are also listed in table \ref{tab:cores} and \ref{tab:area}.}
%Note that there is an additional factor of 3 should be corrected when deriving $\lambda$ from $\lambda_{\rm obs}$ due to fact that 
The derived $\lambda_{\rm obs}$s (shown in table \ref{tab:cores} and \ref{tab:area}) of MM1 and MM3 are all about 1 , suggesting super-critical, on both the core-scale and the clump-scale. 
MM2 is slightly sub-critical ($\sim$0.5) on both scales, which is consistent with the estimate from the pressure and energy density terms (section \ref{sec:pressure_terms}). 
We note that this segregation between MM1/MM3 and MM2 is also consistently seen in $\Sigma_{\rm B}$ which shows the largest values for MM2 on both scales (large values in $\Sigma_{\rm B}$ -- field-to-gravity force ratio -- are equivalent to 
small values in $\lambda_{\rm obs}$ -- mass-to-flux ratio). 
The fact that the averaged $\Sigma_{\rm B}$ remains smaller than one for MM2 on both the core-scale and the clump-scale is due to the large number of positions with very low $\Sigma_{\rm B}$ values in the averaging process. It is, however, evident from figure \ref{Fig:sigmaB} that MM2 displays systematic locations where $\Sigma_{\rm B} $ is clearly larger than one. 

The relative importance of the kinetic support against gravity can be estimated using the virial parameter, 
\begin{equation}
\alpha_{\rm vir} \equiv  \frac{5R}{{\rm G}M}\cdot (\triangle v)^2, 
\end{equation}
%
%
%{\bf where the core mass and radius are defined using the dendrogram analysis (see table \ref{tab:cores_def}). }
where $R$, $M$, $\triangle v$ and G are radius, mass, velocity dispersion and gravitational constant, respectively.
%The derived $\alpha_{\rm vir}$ increases from the clump-scale to the core-scale but still remains $\leq$1.}
It has been suggested that the critical
$\alpha_{\rm vir}$ is $\sim$
$\sim 2$ for non-magnetized clouds \citep{2013Kauffmann}.
Our derived $\alpha_{\rm vir}$ are all smaller than 2, suggesting that these three cores are super-critical if there is no other supporting mechanism.
When taking into account the B field, the $\alpha_{\rm vir, B}$ can be calculated following \citet{1992Bertoldi,2011Pillai} via 
\begin{equation}
    \alpha_{\rm vir, B} = \frac{5R}{GM}\cdot(\triangle v^2 + \frac{\sigma_A^{2}}{6}),
\end{equation}
where $\sigma_A$ is the Alfv\'{e}n velocity for a given magnetic field strength, calculated as $\sigma_A = B/\sqrt{4\pi\rho}$. Here, $B$ is the total B field strength estimated in section \ref{sec:DCF_B}, and $\rho$ is the mass density.
After taking into account the B field contribution, the virial parameter $\alpha_{\rm vir, B}$ is about 5 for the core scale and larger than 2 for the clump scale toward MM2, suggesting that the B field can significantly help to support the MM2 region from collapsing on both scales. This is consistent with the estimate and trend for $\lambda_{\rm obs}$ in the previous paragraph where the MM2 region revealed the smallest values, clearly smaller than one, compared to the other regions that displayed larger values around or above one. We note that the uncertainties in both $\alpha_{\rm vir, B}$ and $\lambda_{\rm obs}$ are non-negligible, but the three regions nevertheless reveal differences and trends that still remain in the presence of these uncertainties.
%%It is evident that the B field toward the MM2 core  does provide \textbf{non-negligible} support against global collapse at the core scale.

%
%
%
%
%section 4.2.6
%
\subsubsection{Robustness and Uncertainties}
\label{sec:robustness}
%The main challenge in comparing and determining the validity of the various approaches resides linking projected observables to 3 dimensional quantities.
A main challenge in computing parameters for the various approaches,  
comparing them and controlling uncertainties (tables \ref{tab:cores} and \ref{tab:area}), are the intrinsic differences in how these parameters are derived. 
The mass $M$ is an {\it integrated} quantity based on a dendrogram which defines a region with a radius $R$ for the core scale, for example.
The magnetic field dispersion 
$\triangle \phi_{\rm B}$ and the field strength $B$ are {\it statistical} quantities, derived from a selected 
ensemble, where single local values for $\triangle \phi_{\rm B}$ and $B$
are not defined.
The force ratio ${\rm \Sigma_B}$ is an {\it averaged} quantity, determined over a selected ensemble of individual ${\rm \Sigma_B}$ values,  where each value is locally defined. 
This means that selecting an area where a quantity is evaluated can already lead to different estimates. 
This could lead to a systematic uncertainty, due to the selected area, which is not accounted for in the statistical error estimates.
We have attempted to control this possible systematic uncertainty by evaluating quantities both on a "core"scale and on a "clump" scale (see the second paragraph of section \ref{sec:GBT} for the definition), whenever the data and technique allowed for this. 
%This possible uncertainty is systematic, i.e., there can be a selection effect,  next to the statistical uncertainties  that directly result from an uncertainty in a measured value that is then further propagated to a final estimate.
In the following we discuss various uncertainties and their possible impact on our estimate of the relative significance between B-field, turbulence, and gravity. 

{\it Selected area.} Table \ref{tab:cores}  and \ref{tab:area} illustrate our findings for the core scale -- including only data limited to the core regions as extracted from the dendrogram analysis (marked as contours in figure \ref{Fig:B_allscale}b and c) -- and  for the larger clump scale  -- including all the available data over the map sizes of figure \ref{Fig:B_allscale}b and c where the observations yielded detections. 
We note that for the large-scale analysis, the separation between MM1 and MM2 is taken to be the middle in between their peaks at a declination of $1\degr 25\arcmin10\arcsec$. 
For the polarization detections, while for MM3 the number of data points for clump and core scale is almost identical,  for both MM1 and MM2 the limitation to the core area substantially reduces the data points. Consequently, there is no measurable difference in the mean value of the
B-field dispersion $\Delta\phi_{\rm B}$ in MM3 from clump to core scale. 
For both MM1 and MM2 the change in $\Delta\phi_{\rm B}$ is within 1$\degr$. However, the statistical uncertainty, driven by the sample size of any of these statistical quantities, increases from about $\pm0.5\degr$ to $\pm2.0\degr$ for both MM1 and MM2 from large to core scale. For MM3, this uncertainty is equally large on either scale with $\pm5.0\degr$.
For the turbulent-to-mean field ratio $\sqrt{\langle B_t^2\rangle^{1/2}/B_0}$, derived from the fitted intercept in figure \ref{Fig:plot_sf}, the choice of clump or core area leads to differences of 1\%, 4\%, 
and 3\% for MM1, MM2, and MM3 with statistical uncertainties between 1\% and 3\%.
Given the mean values in this ratio (around 10\% to 35\%), these uncertainties are a priori not immediately negligible, but noting that the ratio is clearly smaller than one, they do not alter that result. It is also worth noticing that for any of these quantities, their relative order for
the three cores seems to be unchanged when moving from clump to core area, e.g., 
the field dispersion $\Delta\phi_{\rm B}$ is smallest in MM2, followed by MM1 and then MM3, and 
this is seen both for the large area as well as for the limited core area. 
This indicates that our derived estimates related to polarization are robust with respect to the size of the selected area.

Unlike the above polarization-only numbers, parameters that are derived from a combination of polarization and molecular line measurements tend to show clear noticeable differences between the clump and core area (tables \ref{tab:cores} and \ref{tab:area}). The plane-of-sky B field strength $B_{\bot}$ grows by about a factor of three from clump to core scale, while the difference in strengths among the three regions can be up to about a factor of six. Similarly, magnetic field and turbulent pressure ($P_{\rm B}$, $P_{\rm T}$),  and gravitational energy density ($u_{\rm G}$) show a clear increase towards smaller scale by a factor up to about 10 (left panels, figure \ref{Fig:energ}). It is worth noting that for any of these quantities, their relative
ordering among the three regions MM1, MM2, and MM3 seems to be unchanged for both the larger clump and the smaller core region (e.g., the field strength is largest in MM2 and smallest in MM3 with MM1 in between, for both the clump and core scale). This provides additional evidence for 
the robustness of our results, indicating that 
{\it relative trends} can still be captured in the presence of uncertainties and possibly biased selection of areas. Nevertheless, it is central to notice that the {\it change in ratios over scale} among the various constituents  
seems to hold the key for understanding how fragmentation proceeds on 
a next smaller scale (right panels in figure \ref{Fig:energ}; section \ref{sec:fragmentation}). 

The two independently derived parameters dealing with gravity versus magnetic field -- the mass-to-flux ration $\lambda_{\rm obs}$ and the field-to-gravitational force ratio $\Sigma_{\rm B}$ -- show consistent results, i.e., a relative increase in $\lambda_{\rm obs}$ is reflected by a relative drop in $\Sigma_{\rm B}$. This consistent trend is worth noting because $\lambda_{\rm obs}$ is calculated from $B_{\rm total}$ which is corrected with a statistical factor of two for the unknown field orientation with respect to the line of sight while no correction is needed for $\Sigma_{\rm B}$ (see below). Changes between the two scales are rather small for both parameters. All regions with $\lambda_{\rm obs}$ larger than one show values in $\Sigma_{\rm B}$ that are smaller than one. The smaller-than-one values ($\sim 0.5$) in $\lambda_{\rm obs}$ for MM2 are correctly reflected by the largest $\Sigma_{\rm B}$ values ($\sim 0.7 - 0.9$). The fact that the latter one is not larger than one is because $\Sigma_{\rm B}$ is averaged over local $\Sigma_{\rm B}$ values where many are small. Figure \ref{Fig:sigmaB} demonstrates that $\Sigma_{\rm B}$ clearly identifies locations with $\Sigma_{\rm B} > 1$.

Both virial parameters, $\alpha_{\rm vir}$ and $\alpha_{\rm B,vir}$, grow from clump to core scale for all three regions. In this case, the selected area can be crucial as some of the values fall above or below the critical virial threshold depending on the selected area. 

{\it Inclusion of polarization data in the range of 2 to 3$\sigma_{\rm P}$.} We have probed the impact of including data in the range between 2 to $3\sigma_{\rm P}$ as compared to only working with data above $3\sigma_{\rm P}$. For the clump areas in MM1 and MM2, including only data above $3\sigma_{\rm P}$
leads to a field dispersion $\Delta\phi_{\rm B}$ reduced by 3$\degr$ and 1$\degr$, respectively.
The magnetic field strengths $B_{\bot}$ consequently increases by 0.1 and 0.2 mG for MM1 and MM2, respectively.
The turbulent-to-mean field estimates drop by 4\% and 2\%. 
Statistical uncertainties remain almost identical to the full data sets because the 
ensembles of data points are still large. 
Both core areas in MM1 and MM2 contain mostly data above $3\sigma_{\rm P}$. 
In MM3, only five detections are above 3$\sigma_{\rm P}$ which does not allow for a meaningful statistical analysis. 
It is important again to note that -- similar to the above discussed impact of the chosen area -- both including and discarding the 2 to $3\sigma_{\rm P}$ polarization data 
seems to preserve the relative importance among the three cores for all quantities. 

{\it Unknown orientation of the magnetic field with respect to the line of sight.}
The magnetic field-to-gravity force ratio $\Sigma_{\rm B}$ is minimally or not at all affected by the unknown inclination of a magnetic field orientation. 
This is because $\Sigma_{\rm B}$
is the ratio of two angles, where both angles are affected by an inclination correction \citep{2012Koch_a}. 
For the total B field strength when using the DCF method, as we only probe the plane-of-sky B field ($B_{\bot}$), we have adopted a factor of 2, $B_{\rm total}$ = $2\cdot B_{\bot}$,  for a statistical correction that assumes a large number of clouds with B field orientations randomly uniformly oriented with respect to the line of sight \citep{2004Crutcher}. 
This statistical correction with a factor of 2 effectively corresponds to a field orientation of 60$\degr$ with respect to the plane of sky. 
The $B_{\rm total}$ for a single given object, looking at the range of possible orientations, can be about 1 to 3 times the value of $B_{\bot}$ depending on the actual inclination of the magnetic field. This remaining unknown beyond the statistical factor of 2 correction can cause an additional systematic uncertainty.

{\it Cautionary note.} Our comparison of evaluating quantities on a clump area or limited core area and the comparison of discarding or including the 2-to-3$\sigma_{\rm P}$ polarization data reveals some systematic differences that result from these possible selection effects.
These differences are found to be small for the polarization-related quantities, preserving the relative trends in the estimates for the three cores MM1, MM2, and MM3.
While these findings provide good evidence for the robustness of our joint analysis, they 
nevertheless also point at possible subtleties. How significantly the above discussed 
uncertainties can impact final results, will ultimately depend on the exact  
B-field morphology in a source. 
From the numbers presented in the tables \ref{tab:cores} and \ref{tab:area} it is evident that a single parameter
extracted from a single isolated area is only of limited use and any conclusion for an entire region or source based on  such a single parameter can be very misleading.
It is only the comparison among both the different parameters as well as their development over scales that can properly assess the roles of gravity, 
magnetic field, and turbulence in the core formation and its subsequent fragmentation process. 

%{\bf IS THE FOLLOWING STATEMENT NEEDED HERE?} However, the relative significance between 
%magnetic field, turbulence, and gravity for the three cores are found to be different between these two scales. 
%Gravity is found to be dominant at the core scales, while B field dominates toward the clump scale toward MM2.

%
%
%figure 10: illustration
%
%
\begin{figure}[tbh!]
\includegraphics[trim={0 0 0 0},clip,width=0.5\textwidth]{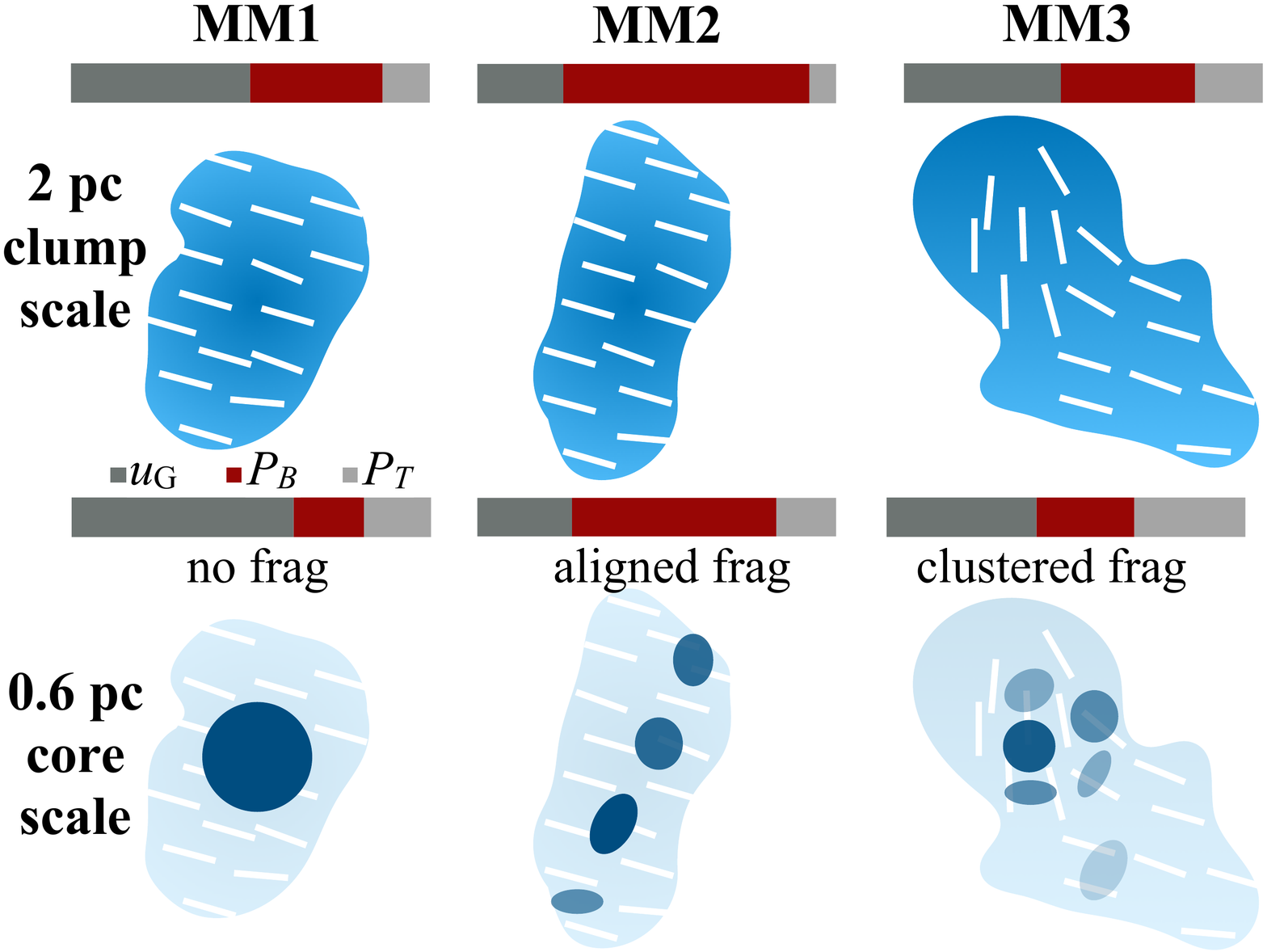}
\caption{
Schematic illustration of the change of the relative significance between $u_{\rm G}$, $P_{\rm B}$, and $P_{\rm T}$ from the 2 pc clump scale (upper panels) to the 0.6 pc core scale (lower panels) for MM1, MM2, and MM3. 
White segments indicate B field orientations.
The bars represent the relative significance between $u_{\rm G}$ (dark grey), $P_{\rm B}$ (red), and $P_{\rm T}$ (light grey), arbitrarily normalized to one, as taken from the Tables \ref{tab:cores} and \ref{tab:area} ignoring their uncertainties. 
See figure \ref{Fig:energ} for the uncertainties. 
The different types of fragmentation (MM1: no fragmentation; MM2: aligned fragmentation; MM3: clustered fragmentation) are suggested to be the result of a different relative importance among the three constituents gravity, B field, and turbulence. This relative importance additionaly seems to evolve differently from the clump to the core scale (figure \ref{Fig:energ}).
%Skematic picture of the structures and the energy balance at the 2 pc clump scale and the 0.6 pc core scale.
}
\label{Fig:illustration}
\end{figure}

%
%
%
%
%
%\subsection{Zooming in onto Smaller Scales: Different Types of Fragmentation and Gravity -- B-field -- Turbulence Interplay
%\label{sec:fragmentation}}
\subsection{Zooming in onto Smaller Scales
\label{sec:fragmentation}}

\subsubsection{Different Types of Fragmentation and Gravity -- B-field -- Turbulence Interplay}

In the previous sections we have presented
observational facts that characterize the region
from the outer filamentary zones to the inner clump/core regions,
namely (1) a mostly uniform large-scale B field perpendicular to the filament is observed towards MM1 and MM2, while a bending B field is seen closely aligning with the MM3 major axis; 
(2) the velocity gradients are closely aligned with the B field toward MM1 and MM2, while they show systematically larger misalignements in MM3;
(3) different values in the dispersion of the B field orientations are observed, being
smallest in MM2, more than twice as large in MM3, and intermediate in MM1.
Additionally, estimating the magnetic field (B), turbulence (T), and gravity (G) with various
techniques has led us to conclude
that ${\rm G}>{\rm B}\sim{\rm T}$ in MM1, ${\rm B}\geq{\rm G}>{\rm T}$ in MM2, and ${\rm G}>{\rm T}\sim{\rm B}$ in MM3 on the 0.6 pc core scale.
This relative importance among B, T, and G has evolved from and changed with respect to the larger 2 pc clump scale
(tables \ref{tab:cores}, \ref{tab:area}) --
at the clump scale, B is dominant in MM2, while G, B and T appear to be comparable to each other in MM3, and G and B are about similar in MM1.
Moreover, this change has developed differently for the three cores MM1, MM2, and MM3, as we argue in the paragraphs below (see also figure \ref{Fig:energ}).

{\it Are these clump-scale and core-scale observational clues and estimates holding the key to predicting or explaining fragmentation on the next smaller scale?}
Observations with the SMA \citep{2014Zhang} and CARMA \citep{2014Hull} have already resolved smaller-scale features in the MM1, MM2, and MM3 cores in G34. 
We propose that three different types of fragmentation are present (panels d, e, f in figure \ref{Fig:B_allscale}): {\it (i) clustered fragmentation in MM3} with B field
orientations differing by up to $90^{\circ}$ (though coverage is incomplete) and fragments being distributed and scattered around in the original clump volume. -- Qualitatively, this is expected to happen only if the B field is not dominant, such that turbulence can lead to scattered small seeds that can locally collapse; 
{\it (ii) aligned fragmentation in MM2} with B-field orientations remaining parallel
to the larger-scale field and fragments lining up along a direction that is approximately perpendicular to the magnetic field orientation. -- Qualitatively, this occurs only for a strong B field
that dominates over turbulence and where local collapse mostly happens along field lines; {\it (iii) no fragmentation in MM1} with field orientations displaying systematic deviations and bending in the north that is likely due to a gravity-driven dragged-in motion,
preserving one single intact core. -- Qualitatively, this is possible if gravity dominates over
both B field and turbulence enabling a global collapse. 

\subsubsection{Change in Relative Significance over Scale: Subtle Balance among Gravity, B field, and Turbulence}

It is instructive to look at the ratios between the gravitational energy density $u_{\rm G}$ and the magnetic and turbulent pressure $P_{\rm B}$ and $P_{\rm T}$ (figure \ref{Fig:energ}, extracted from tables \ref{tab:cores} and \ref{tab:area}) to more quantitatively understand the above proposed fragmentation scenario. We first note that $u_{\rm G}$, $P_{\rm B}$, and $P_{\rm T}$ clearly grow from the larger 2 pc clump to the smaller 0.6 pc core scale, as can be seen in the three left panels of the figure. This is not surprising as such because $u_{\rm G}$ obviously grows towards denser regions, the field strength and hence $P_{\rm B}$ grows with density if flux-freezing is valid, and $P_{\rm T}$ scales with the obviously growing density while the velocity dispersions are measured to change only little.  It is, nevertheless, remarkable that each one of the three quantities shows a very similar increase (i.e., slope) in each of the three regions MM1, MM2, and MM3 (see the 9 curves in the three left panels in figure \ref{Fig:energ}.)
{\it With such similar trends from clump to core scale, can we tell why fragmentation still develops so differently?}
For this purpose, turn now to the ratios between the quantities, i.e., the three right panels in figure \ref{Fig:energ}.  
Now we can see variations in slope among the 9 curves shown, thereby exploring more subtle effects.  
While not all of these variations are larger than the associated uncertainties, some intriguing correlations are seen.  
For example, the MM1 region, that shows no fragmentation, has the highest $u_{\rm G}$/$P_{\rm T}$ and the highest $u_{\rm G}$/$P_{\rm B}$, especially at the core scale.  
The MM2 region, that shows aligned fragmentation with fragments running along a line perpendicular to the B-field at 2 pc scale, has consistently higher $P_{\rm B}$/$P_{\rm T}$ and consistently lower $u_{\rm G}$/$P_{\rm B}$, as compared to the other regions, with all regions showing a decrease in $P_{\rm B}$/$P_{\rm T}$ in moving from the clump scale to the core scale.
%\textbf{Unlike the evolution of $u_{\rm G}$, $P_{\rm B}$, and $P_{\rm T}$ (left panels of figure 9), the ratios of these quantities (right panels of figure 9) show apparent differences for the MM1, MM2, and MM3 regions.  
%For $P_{\rm B}/P_{\rm T}$, there is a hint of differences in the ratio quantities (for example, ratio in MM2 evolves from 9 at the clump scale to 3 at the core scale). However, such variations are within the uncertainty.
%For  $u_{\rm G}$/$P_{\rm T}$ and $u_{\rm G}$/$P_{\rm B}$, it is evident that the differences are both in slope and in quantity, especially the ratio $u_{\rm G}$/$P_{\rm T}$.
Accordingly, we propose that it may be the {\it change in the relative significance among the three constituents gravity, B-field, and turbulence from clump to core scale that sets the initial conditions for fragmentation.}  
I.e., since the overall evolution of each of $u_{\rm G}$, $P_{\rm B}$, and $P_{\rm T}$ is similar, we suggest that the resulting differences in fragmentation type among the three regions might actually be the consequence of a rather subtle balance among these three ingredients, as illustrated in figure \ref{Fig:illustration}.

\section{Conclusion}
We present the results of the dust polarization observations at 350$\mu$m toward the infrared dark cloud G34.43 and the analysis on the correlation between the local velocity gradient and the local B field.
We find a close alignment between local B field orientations and local velocity gradients toward the MM1/MM2 ridge, suggesting gas motions are influenced by the magnetic field morphology or vice versa.
%We derive all the physical parameters based on these results. 
We apply and compare various techniques available in the literature to estimate the B field strength and the relative importance between gravity (G), magnetic field (B) and turbulence (T).
While the different techniques essentially lead to consistent results (summarized with the last column in table \ref{tab:cores} and \ref{tab:area}),
solely focusing on one technique might lead to an incomplete or less clear picture about the relative importance between gravity, magnetic field, and turbulence. 
We note important advantages, shortcomings, and differences in section \ref{sec:GBT}.
As the derived parameters can be sensitive to the selected area (see the discussion in section \ref{sec:robustness}), we consider two different representative scales, namely the smaller core area at a scale of 0.6 pc and the larger clump area at a scale of 2 pc.

We find that ${\rm G}>{\rm B}\sim{\rm T}$ in MM1, ${\rm B}\geq{\rm G}>{\rm T}$ in MM2, and ${\rm G}>{\rm T}\sim{\rm B}$ in MM3 on the 0.6 pc core scale.
This relative importance seems to have evolved from and changed with respect to the larger 2 pc clump scale (tables \ref{tab:cores} and \ref{tab:area}, figure \ref{Fig:energ}).
We further look at the ratios between the gravitational energy density $u_{\rm G}$ and the magnetic and turbulent pressure $P_{\rm B}$ and $P_{\rm T}$ (figure \ref{Fig:energ}) to more quantitatively understand the different fragmentation scenarios at 0.1 pc scale shown in figure \ref{Fig:B_allscale}d,e,f (section \ref{sec:fragmentation}). 
The estimated $u_{\rm G}$, $P_{\rm B}$, and $P_{\rm T}$ clearly grow from the larger 2 pc clump to the smaller 0.6 pc core scale with a similar trend. 
Interestingly, we find that the ratios among them (right panels in figure \ref{Fig:energ}) show clear differences for the MM1, MM2, and MM3 regions, although the uncertainties of the derived $u_{\rm G}$, $P_{\rm B}$ and $P_{\rm T}$ are non-negligible. 
We propose that it might be the change in the relative significance among the three constituents gravity, B field, and turbulence from clump to core scale that sets the initial conditions for fragmentation.

\acknowledgments
We are grateful to the CSO staff for very efficient support, both on site as well as during the later remote observations. Y-WT is supported by the Ministry of Science and Technology (MoST) in Taiwan through grant MoST 108-2112-M-001-004-MY2. PMK acknowledges support from
MoST 107-2119-M-001-023
and from an Academia Sinica Career Development Award.
ADC acknowledges the support from the UK STFC consolidated grant ST/N000706/1.

{\it Facilities:} \facility{SHARP}, \facility{CSO}.

%
%
% Figures
%

%
%
%
%\begin{thebibliography}{}
%\bibitem[Auri\`ere(1982)]{aur82} Auri\`ere, M.  1982, \aap,
%    109, 301
%\end{thebibliography}
\bibliographystyle{apj}
\bibliography{G34}

%\clearpage

%
\end{document}